\documentclass[]{solarphysics}
\usepackage{sunback_preamble}
\usepackage{wrapfig}
\usepackage[dvipsnames]{xcolor}
\usepackage{graphicx}
\usepackage{tabularx}
\usepackage{lscape}
\usepackage{booktabs}
\usepackage{caption}
\usepackage{listings}
\usepackage[normalem]{ulem}  % for strikethrough
\usepackage{rotating}

\usepackage{makecell}    % for nicer multi-line entries
\usepackage[margin=0.85in, bottom=1.5in, top=0.85in]{geometry}

\newcommand{\changed}[1]{#1}
\newcommand{\changedeq}[1]{#1}
\newcommand{\remove}[1]{}

\usepackage[useregional]{datetime2}
\newcommand{\usdate}{\DTMdisplaydate{\the\year}{\the\month}{\the\day}{-1}}

\makeatletter
\renewcommand{\idline}{%
  \if@noid\else
    \stepcounter{outputpage}%
    \vspace*{-50pt}% tighten vertical space above the idline
    \rlap{\smash{\vtop to 0pt{%
      \vfil\hbox to\textwidth{%
      \hfil\footnotesize\tt
      SOLA: \jobname.tex; \usdate;~\timenow;~p.\theoutputpage}}}}%
  \fi
}
\makeatother

\begin{document}
\begin{opening}
    % \title{Improving Visualization of the Solar Corona with the Radial Histogram Equalizing Filter}
    \title{Visualization of High Dynamic Range Solar Imagery and the Radial Histogram Equalizing Filter}
    \author[addressref={swri,lasp},corref,email={Chris.Gilly@colorado.edu}]{\inits{C.R.}\fnm{C.R.}~\lnm{Gilly}\orcid{0000-0003-0021-9056}}
    \author[addressref={lasp},email={Steven.Cranmer@lasp.colorado.edu}]{\inits{S.R.}\fnm{S.R.}~\lnm{Cranmer}\orcid{0000-0002-3699-3134}}

    \runningauthor{Gilly and Cranmer}
    \runningtitle{Visualization of HDR Solar Imagery with RHEF}
    \address[id=swri]{Department of Solar and Heliospheric Physics, Southwest Research Institute, Boulder, CO 80302, USA}
    \address[id=lasp]{Department of Astrophysical and Planetary Sciences, Laboratory for Atmospheric and Space Physics, University of Colorado Boulder, CO 80303, USA}

    \begin{abstract}
        Standard visualizations of Extreme Ultraviolet (EUV) solar imagery often fail to convey the full complexity of the Sun's corona, especially in faint off-limb regions. This can leave the misleading impression of the Sun as a bright ball in a dark void, rather than revealing it as the dynamic, structured source of the solar wind and space weather. A variety of enhancement algorithms have been developed to address this challenge, each with its own strengths and tradeoffs.
        We introduce the Radial Histogram Equalizing Filter (RHEF), a novel hybrid technique that optimizes contrast in high dynamic range solar images. By combining the spatial awareness of radial graded filters with the perceptual benefits of histogram equalization, RHEF reveals faint coronal structures and works out of the box---without requiring careful parameter tuning or prior dataset characterization. RHEF operates independently on each frame, and it enhances on-disk and off-limb features uniformly across the field of view. For additional control, we also present the Upsilon redistribution function---a symmetrized cousin of gamma correction---as an optional post-processing step that provides intuitive programmatic tonal compression.
        We benchmark RHEF against established methods and offer guidance on filter selection across various applications, with examples from multiple solar instruments provided in an appendix. Implemented and available in both Python \texttt{sunkit\_image} and IDL, RHEF enables immediate improvements in solar coronal visualization.
    \end{abstract}

\end{opening}
%\clearpage

\begin{figure}[ht!]
    \includegraphics[width=\linewidth]{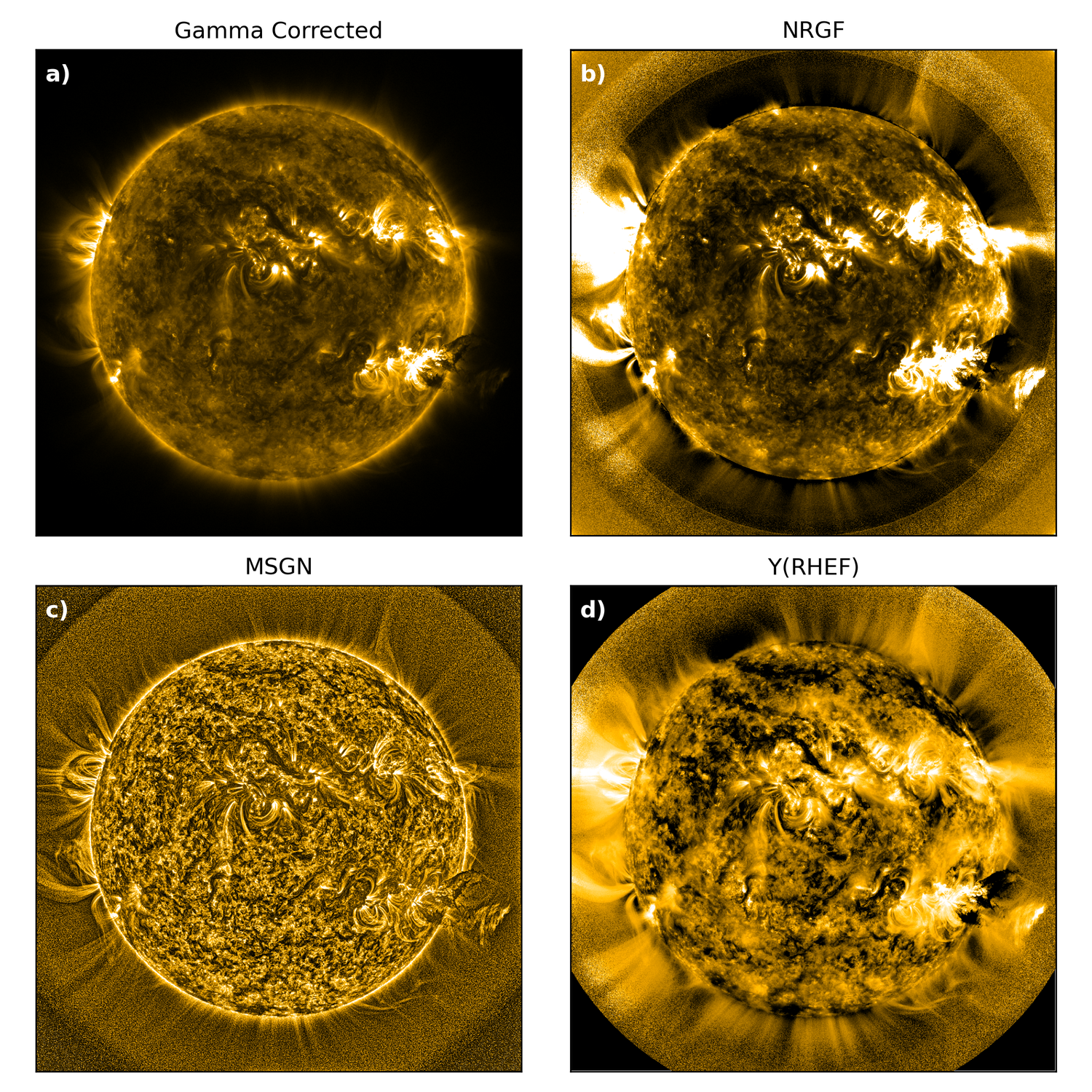}
    \caption{Comparison of contrast enhancement techniques applied to the SunPy AIA 171~\AA{} sample image. The top-left panel shows the image with basic gamma correction (i.e., without high dynamic range (HDR) enhancement). The top-right and bottom-left panels show results from the Normalizing Radial Graded Filter (NRGF) and Multiscale Gaussian Normalization (MSGN), respectively (we use these for comparison, but note that they were never designed to be used on-disk). The bottom-right panel displays the result of the Radial Histogram Equalizing Filter (RHEF) and Upsilon \(\Upsilon\), which preserves detail across both on-disk and off-limb regions while enhancing the visibility of faint coronal structures.}\label{fig:familyphoto}
\end{figure}
\vfill
\section{Introduction} \label{sec:intro}

Visualizing images of the solar corona is a fundamental task in heliophysical research, with direct implications for understanding solar dynamics and forecasting space weather. However, this task presents persistent challenges due to the wide range of radiance levels across many Sun-centered fields of view. In Extreme Ultraviolet (EUV) observations, two distinct forms of dynamic range complexity emerge.

First, the steep radial falloff in emission---driven by decreasing plasma density and emissivity with height---causes large-scale coronal structures to fade rapidly with distance from the solar surface. These structures often become difficult to visualize without specialized processing \citep{DelZanna2018}. Second, embedded within these gradients are small-scale, high-frequency features whose contrast is subtle relative to the surrounding background. Although faint, these features are physically meaningful and typically lie above the instrumental noise floor \citep{Aschwanden2010}. To study both classes of features---especially those extending into the off-limb corona---advanced image processing is essential. Such methods compress the dynamic range to enhance visibility, ideally preserving structural fidelity without discarding small-amplitude transients of interest.

A variety of algorithms have been developed to address these challenges. Solar image processing has evolved to include several families of enhancement techniques that fall broadly into three categories: Radial Gradient Filtering (RGF), Histogram Equalization (HE), and Multiscale Methods (MSM)---each offering distinct advantages for interpreting high dynamic range (HDR) solar imagery \citep{Morgan2007, Morgan2014}. However, each approach has limitations, and the choice among them depends heavily on the observer's goals and the structures of interest.

As the demand grows for tools capable of detecting low-contrast features and tracking dynamics across a broad range of spatial scales, such filters have become indispensable in both research and operational space weather contexts \citep{West2022}. We introduce a new hybrid algorithm: the Radial Histogram Equalizing Filter (RHEF). This method specifically targets the shortcomings of existing techniques, offering an efficient, parameter-free solution that enhances both large-scale coronal architecture and fine-scale detail.

This article is organized as follows. Section~\ref{sec:background} provides a brief overview of solar image preparation and processing, including a discussion of their downstream implications for visualization in Section~\ref{sec:implications}. Section~\ref{sec:review} reviews three major families of HDR compression techniques. Section~\ref{sec:algorithms} presents two new tools: the RHEF and the Upsilon redistribution function (\(\Upsilon\)). In Section~\ref{sec:results}, we benchmark these tools against established methods and demonstrate their performance across representative features. We conclude in Section~\ref{sec:conclusion} with a summary of key findings and directions for future work.

Appendix~\ref{sec:appendix_instruments} showcases the application of RHEF to a range of solar instruments. Appendix~\ref{sec:appendix_synoptic} introduces a new FIDO Client for downloading reduced-resolution AIA imagery from the synoptic series (1024x1024 at a 2-minute cadence), and Appendix \ref{sec:appendix_python} gives a brief python RHEF usage example.
% , and Appendix \ref{sec:appendix_IDL} provides an IDL implementation.

\section{Image Processing Primer}\label{sec:background}
Raw telescope data must undergo essential pre-processing before it becomes suitable for scientific analysis. Using the Solar Dynamics Observatory's Atmospheric Imaging Assembly (SDO/AIA; \citealt{Lemen2012}) as a representative example, we outline common image corrections in solar physics, noting that specific procedures vary between instruments. For those new to this workflow, Python-based tools such as \texttt{SunPy} \citep{Barnes2020} provide a solid foundation for data handling. More detailed descriptions of standard image reduction steps are available in \citet{Aschwanden2010}, \citet{Boerner2012a}, and \citet{Possel2020}.

\subsection{Implications of Filtering}\label{sec:implications}

While calibrated images are essential for quantitative science, they are often suboptimal for direct visualization. The solar corona presents an extreme dynamic range that challenges standard display techniques, with bright on-disk emission and faint off-limb structures coexisting in the same frame. As a result, nonlinear filtering algorithms are commonly used to compress this range, enhancing visual clarity and feature detectability.

However, these filters---such as those listed in Table~\ref{tab:Methods}---modify pixel values in nonlinear, spatially varying ways. Because they do not apply uniform scaling, filtered intensities lose their original linear relationship to physical units. This compromises the quantitative accuracy of the image, rendering it unsuitable for analyses that rely on calibrated radiance---such as plume/interplume contrast, emission measure, or temperature and density diagnostics \citep[e.g.,][]{Boerner2012a}.

\begin{table}[ht!]

    \begin{tabular}{lll}
        \toprule
        \textbf{Name} & \textbf{Abbrev} & \textbf{Reference} \\
        \midrule
        \multicolumn{3}{l}{\textbf{Radial Graded Filters (RGF)}}\\
        Normalizing Radial Graded Filter         & *NRGF    & \cite{Morgan2006} \\
        AIA\_RFILTER                               & -        & McCauley, Cranmer, and Engell (2010) \\
        AIA\_OFFLIMB                             & -        & McCauley, Cranmer, and Engell (2010) \\
        Fourier Norm. Radial Graded Filter       & *FNRGF   & \cite{Druckmullerova2011} \\
        Simple Radial Gradient Filter            & SIRGRAF  & \cite{Patel2022} \\
        SWAP Filter                              & SWAP     & \cite{Seaton2023} \\
        \addlinespace
        \multicolumn{3}{l}{\textbf{Histogram Equalization (HE)}} \\
        Adaptive Histogram Equalization          & AHE      & \cite{Pizer1987} \\
        Contrast-Limited Adaptive HE             & CLAHE    & \cite{Pisano1998} \\
        Adaptive Circular HP Filter              & ACHF     & \cite{Druckmuller2006} \\
        Noise Adaptive Fuzzy Equalization        & NAFE     & \cite{Druckmuller2013} \\
        NAFE (Variable Neighborhood)             & NAFEVN   & \cite{Druckmuller2014} \\
        Radial Histogram Equalizing Filter       & *RHEF    & \cite{Gilly2022} and \textbf{this paper} \\
        \addlinespace
        \multicolumn{3}{l}{\textbf{Multiscale Methods (MSM)}} \\
        Wavelet Transform                        & WT       & \cite{Stenborg2003} \\
        Multi-Scale Gaussian Norm                & *MSGN     & \cite{Morgan2014} \\
        Radial Local Multiscale Filter           & RLMF     & \cite{Qiang2020} \\
        Wavelet Optimized Whitening              & *WOW     & \cite{Auchere2023a} \\
        \bottomrule
    \end{tabular}
    \caption{
        A non-exhaustive overview of filtering methods across several families. Methods marked with an asterisk (*) are available in the SunPy-affiliated package \texttt{sunkit-image}. Note that some hybrid methods may span multiple categories.
    }
    \label{tab:Methods}
\end{table}

Despite this loss of photometric fidelity, filtered images remain valuable for scientific workflows that rely on morphology rather than calibrated intensity. These include segmentation, edge detection, and feature tracking used to identify structures such as plumes, loops, sunspots, and coronal holes. As observational cadence and spatial resolution improve, the ability to reveal faint, dynamic features becomes increasingly critical.

Filtering can also enhance physical interpretation in motion-tracking workflows. For instance, identifying solar wind outflows enables acceleration profile derivation that constrains coronal heating models \citep{Miralles2001,Telloni2019, Cranmer2020c}. Careful spatial filtering can also help identify and track small density enhancements that trace larger-scale outflows and inflows \citep[][]{SheeleyJr.1997, DeForest2014, Lopez-Portela2018} Similarly, dynamic range compression improves visibility of flare morphologies, wavefronts, and CME propagation, aiding in the linkage between inner and outer corona observations \citep{Vourlidas2006, Gilly2022}.

While nonlinear filters alter absolute intensities, they often illuminate physically meaningful structure that would otherwise remain hidden. In the next section, we group these techniques into several families, highlighting their comparative strengths and trade-offs---setting the stage for the new method introduced in Section~\ref{sec:algorithms}.

% \clearpage

\subsection{Review of Filter Families}
\label{sec:review}

Building on the discussion of filtering implications, we categorize the main families of HDR enhancement algorithms used in solar image processing. While many image processing techniques originate from other fields or industry---and may fall outside these categories, including general enhancement algorithms \citep[see, e.g.,][]{Andiani2024} or manual tools like Adobe Photoshop---this review focuses on methods commonly applied in solar physics.

Representative examples of each family are listed in Table~\ref{tab:Methods}, with their effects illustrated in Figure~\ref{fig:familyphoto}, showing different methods applied to the same AIA 171~\AA\ image. For an in-depth review and comparative analysis of some selected filters, see Chapter 4 of \citet{Gilly2022}.

Each technique has made important contributions to solar image processing, but their effectiveness often depends on careful parameter tuning---such as selecting spatial scales, normalization thresholds, or kernel widths---posing challenges for intuitive use. Although highly effective in specific applications, few methods provide a unified approach capable of simultaneously enhancing both the bright solar disk and the faint extended corona.

\subsubsection{Radial Graded Filters (RGFs)}

Radial Graded Filters are a class of algorithms designed specifically to suppress the dominant radial brightness gradient in solar images, especially in off-limb regions. Originally developed for white-light eclipse and coronagraph imagery \citep{Newkirk1967, Eddy1989}, RGFs aim to flatten this large-scale structure, thereby enhancing finer embedded features.

\changed{The most widely used method in this family is the Normalizing Radial Gradient Filter (NRGF), which usually subtracts the \changed{mean} brightness at each radius and normalizes by the standard deviation. Variants of this technique differ in how they estimate the radial profile, including which statistics to use for subtraction and normalization, and whether they use a single image or an ensemble. The NRGF in specific was originally conceived to overcome the problem of the non-radial appearance of streamers in coronagraph images. Calibrated data show that these structures are radial above about \(2.5R_\odot\), but they appeared to narrow with distance due to poor choices of image processing (e.g.~log brightness) \citep{Morgan2006}.}

\changed{RGFs are particularly effective off-limb, where the dominant brightness gradient is radial. A weaker radial dependence also exists just inside the limb due to limb darkening/brightening. Across most of the disk, however, no strong radial gradient is present, so RGFs are not strictly justified there. For simplicity and interpretability, however, RGFs are often applied to the full image. Using annuli on-disk is no worse than other arbitrary choices, and switching to a different scheme would create a visual discontinuity (typically at $r=R_\odot$), producing an artificial ``knee'' or scar that does not correspond to a real solar feature. In practice, this means RGFs are most valid off-limb but are commonly extended across the image for continuity. One caveat is that near disk center the methods suffer from small-number statistics, making the region more sensitive to bright or dark features (See Section~\ref{sec:mitigation} for more details).}

Figure~\ref{fig:familyphoto}(b) shows a sample result from NRGF processing, which has been used extensively in eclipse, coronagraphic, and AIA studies \citep[e.g.,][]{Morgan2007, Habbal2010, Morgan2010, Wang2010, Druckmullerova2011, Morgan2012, Byrne2012, Byrne2014, Alzate2017, Cho2019, Lee2020, Ruminska2022}.

SolarSoft, the IDL-based community software suite \citep{Freeland2004}, includes two legacy implementations in this family: \texttt{AIA\_OFFLIMB} and \texttt{AIA\_RFILTER}, originally developed by A. Engell and S. Cranmer and later implemented by P. McCauley. These tools have supported numerous studies \citep{Ma2011, Kliem2013, Masson2014, Su2015, Mishra2020}.

\subsubsection{Multiscale Methods (MSMs)}

 \changed{Multiscale methods generalize RGFs by decomposing an image into a set of characteristic spatial scales, allowing selective enhancement or suppression of structural features.} Instead of assuming radial symmetry, these algorithms operate fully in two dimensions and work across multiple spatial-frequency bands. This approach is highly flexible and effective but often require careful per-dataset tuning of scale parameters, normalization heuristics, and kernel sizes. They can also introduce spatial artifacts or haloing effects due to overlapping filter kernels---especially near sharp boundaries.

The Multiscale Gaussian Normalization (MGN/MSGN) filter \citep{Morgan2014} is a widely used MSM, built by combining Gaussian-blurred versions of the image at several scales. It requires a few tunable parameters: the Gaussian widths (\texttt{sigma}), a scaling factor (\texttt{k}), and a global gamma correction (\texttt{gamma}, with limits \texttt{gamma\_min} and \texttt{gamma\_max}). Additional weights (\texttt{h}, \texttt{weights}) control the balance between global and multiscale contributions, while kernel truncation (\texttt{truncate}) and clipping (\texttt{clip}) handle computational and numerical stability.

The Wavelet-Optimized Whitening (WOW) algorithm \citep{Auchere2023a} takes a similar approach to MGN, using a wavelet basis to equalize contrast across spatial-frequencies. Fourier filtering is another common strategy, again isolating and selectively enhancing features based on their spatial frequency components. More advanced variants, such as the noise-gating technique of \citet{DeForest2017}, apply Fourier-domain filtering across time series to suppress noise and enhance low-contrast features.

Figure~\ref{fig:familyphoto}(c) shows MSGN applied to AIA data as a representative example. MSMs have been particularly successful in eclipse imagery and high-resolution off-limb studies \citep[see, e.g.,][]{Pant2015, Morton2016, French2019, Okane2019, French2020, Williams2020, Boe2021, Morgan2022, Auchere2023b}.

\subsubsection{Histogram Equalization (HE)}

Histogram equalization methods enhance contrast by remapping image intensities to maximize the output dynamic range---typically by flattening or reshaping the histogram. Unlike gradient- or frequency-based filters, HE algorithms operate directly on intensity distributions, making them particularly effective in high dynamic range contexts with limited perceptual contrast.

Global HE methods compute a cumulative distribution function (CDF) from the full image and apply a global intensity remapping \citep{Hummel1977, Ketcham1976, Pizer1981}. Adaptive variants, such as Adaptive Histogram Equalization (AHE; \citealt{Pizer1987}), divide the image into local neighborhoods for independent equalization. However, AHE can amplify noise in low-signal regions. Contrast-Limited AHE (CLAHE; \citealt{Pisano1998}) addresses this by capping histogram bins to reduce the influence of outliers.

Histogram equalization and percentile-based remapping are closely related, both relying on the image's cumulative distribution. This equivalence is central to the design of the Radial Histogram Equalizing Filter (RHEF), introduced in Section~\ref{sec:algorithms}, which applies a percentile-based transformation within concentric annuli centered on the solar disk. By incorporating geometric context, RHEF avoids the need for user-defined parameters while enhancing structural visibility across both on-disk and off-limb regions. Unlike prior radial filtering approaches such as ACHF or NAFE, which apply histogram manipulation heuristics or fuzzy contrast enhancement, RHEF uses strict percentile remapping within geometrically defined annuli. This simplifies interpretation and reproducibility, avoids heuristic tuning, and provides a consistent visual mapping without amplifying noise in low-signal regimes. Figure~\ref{fig:familyphoto}(d) shows an example application of RHEF, which leverages the benefits of histogram equalization while maintaining radial awareness for solar images.

% \subsubsection{Temporal Background Subtraction (TBS)}
% \label{sec:background_subtraction}

% Temporal background subtraction is a class of enhancement methods that isolate dynamic structures by removing a precomputed background from each image \citep[e.g.,][]{Morrill2006, Thompson2010, Seaton2023}. These approaches are common in coronagraphic and heliospheric imaging, where the slowly evolving large-scale corona must be separated from transient features like CMEs or solar wind streams.

% Unlike annulus-based techniques, which require radial binning, background subtraction inherently accounts for radial intensity falloff but depends on a timeseries to characterize the background. Implementations range from static background models to centered running averages and rolling median or percentile filters over hours to weeks.

% Because TBS relies on multiple frames---often using both past and future data---it enables sensitive change detection but introduces latency, especially with long time windows. This limits its suitability for real-time use or missions with sparse cadence.

% In contrast, filters like RHEF operate on single frames, providing immediate enhancement without temporal context. This makes them ideal for low-latency pipelines, preview modes, and situations where background models are unavailable.

\section{New Algorithms}\label{sec:algorithms}

\subsection{Radial Histogram Equalization (RHE)} \label{sec:rhef}

\begin{figure}[ht!]
    \centering
    \includegraphics[width=0.98\linewidth]{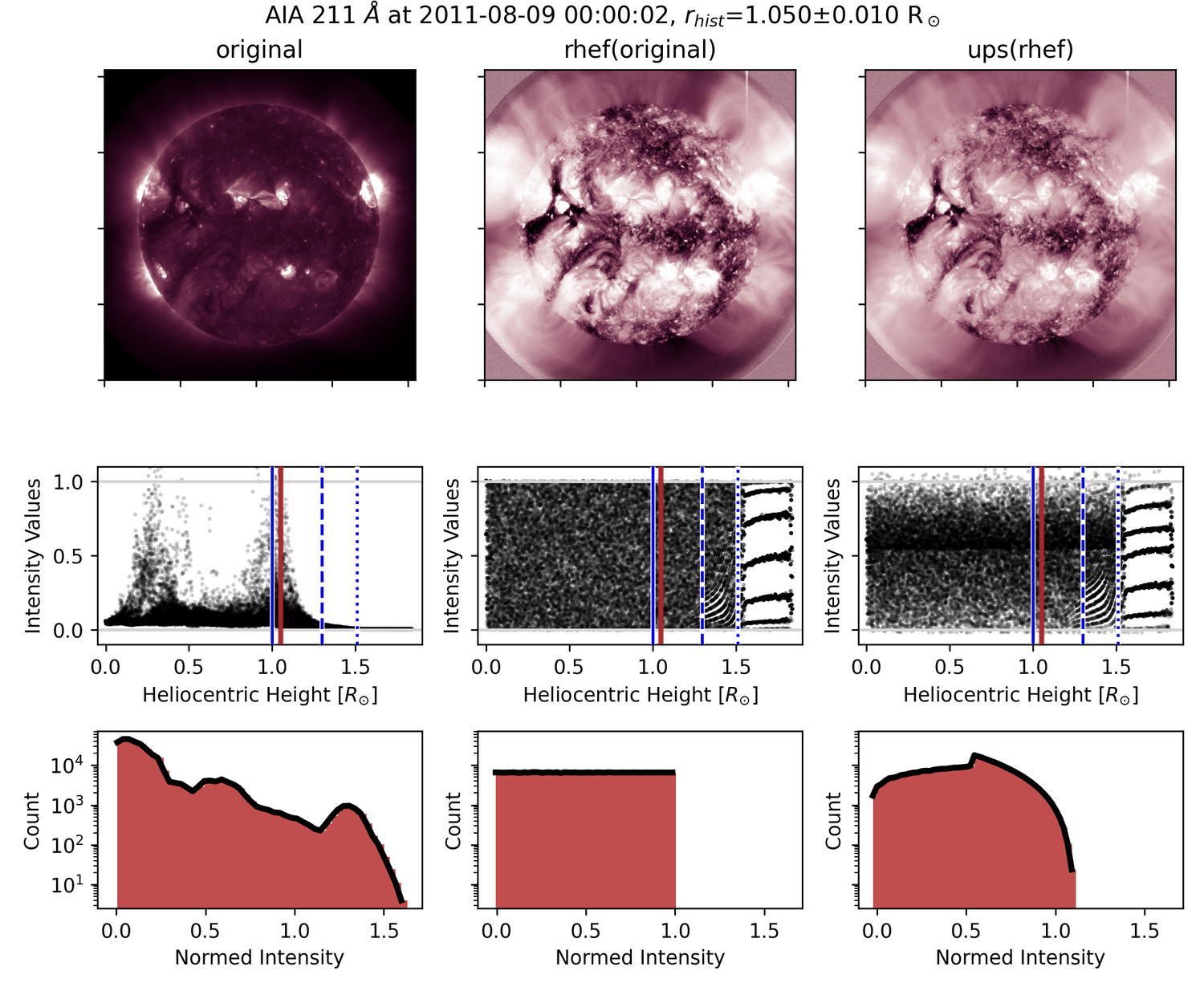}
    \caption{\changed{Effect of the filter on an AIA 211 image. Each column represents a different filter procedure. The top row shows the image, the middle row shows the full histogram of values as a function of height, and the bottom row shows a histogram of values at one height, as indicated by the red vertical bar in the second row. The blue vertical lines at increasing heights show the limb, detector edge, and optical edge respectively. Note that this figure is showing the effect of $\Upsilon_L \neq \Upsilon_H$. Note also that the ``comb-like'' structures in the histogram come from the small number of integer values recorded outside the optical path on the detector.}}\label{fig:rheTriPanel}
\end{figure}

The Radial Histogram Equalizing Filter (RHEF) is a hybrid image-processing technique that enhances high dynamic range (HDR) solar images. It combines the radial spatial awareness of RGF methods with the distribution-flattening benefits of histogram equalization to deliver parameter-free contrast normalization across both on-disk and off-limb regions of an image. Figure~\ref{fig:rheTriPanel} demonstrates the effect of the filter, showing how the unstructured input is transformed by the ranking into an even distribution of intensities bounded between 0 and 1.

Inspired by Radial Gradient Filters (RGFs), RHEF replaces mean or median subtraction with rank-based remapping: each pixel's value is transformed according to its percentile rank within a concentric annulus centered on disk center. This operation enhances contrast without requiring assumptions about the dataset or manual tuning.

Unlike standard Adaptive Histogram Equalization (AHE), which computes local contrast per pixel neighborhood, RHEF uses pseudo-global radial annuli to maintain geometric context. This is particularly advantageous for wide-field solar imagery, where brightness can drop steeply with radius and traditional filters struggle to balance disk and limb visibility. Annulus widths are chosen to be approximately one pixel wide, balancing spatial resolution and statistical stability. Fixed-width radial bins are used, though future enhancements may adopt adaptive schemes to respond to local image structure. If a bright feature dominates a narrow bin, faint ring artifacts can occur (see Section~\ref{sec:overall}). These can be mitigated by tuning bin widths or using overlapping smoothing strategies.

The algorithm is implemented in the \texttt{radial} module of the SunPy-affiliated package \texttt{sunkit\_image}, and is compatible with \texttt{SunPy.map.Map} objects. It uses efficient NumPy and SciPy array operations to compute annular rankings. By default, sorting is performed via \texttt{argsort}, though users may choose \texttt{scipy.stats.rankdata} to handle ties using different heuristics \citep{Virtanen2020}.

For each annulus \( A_i \), pixel intensities are remapped using:
\begin{equation} \label{eq:rhef_rankdata}
I_{\mathrm{out}}[A_i] = \frac{1}{N_{A_i}} \textsf{rank}(I_{\mathrm{in}}[A_i]),
\end{equation}
where \( N_{A_i} \) is the number of pixels in annulus \( A_i \), and \textsf{rank} is the chosen ranking function. The result is a normalized output with a fixed distribution range \([0, 1]\) and consistent visual properties.
% See appendix \ref{sec:appendix_bestpractices} for a minimal usage example. Further
Examples and gallery notebooks are available in the online documentation\footnote{\url{https://docs.sunpy.org/projects/sunkit-image/en/stable/generated/gallery/radial_histogram_equalization.html}}.

\begin{figure}[ht!]
    \centering
    \begin{minipage}{0.55\textwidth}
        \includegraphics[width=\linewidth]{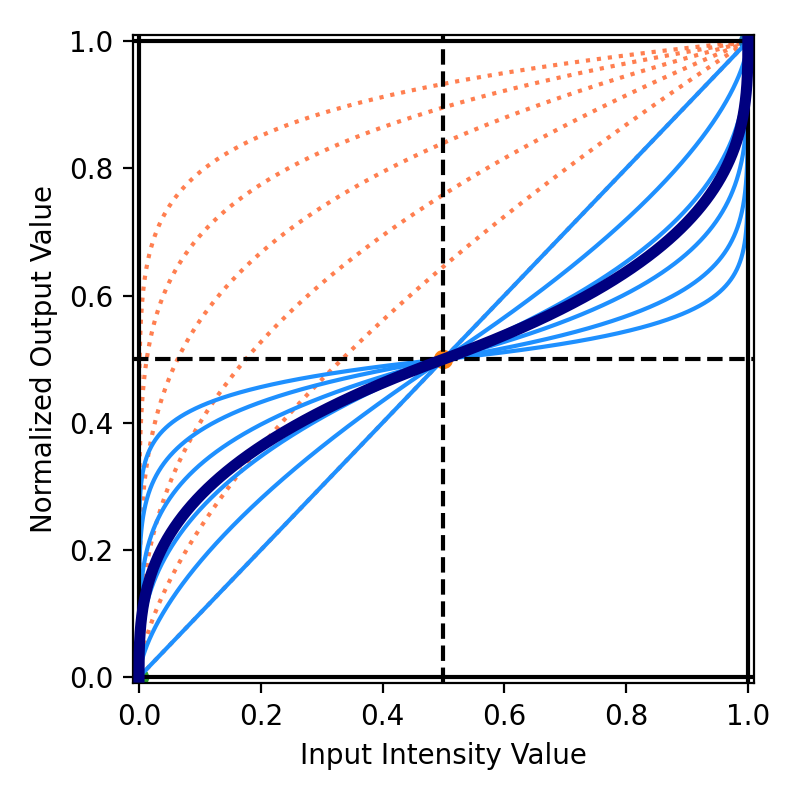}
    \end{minipage}%
    \begin{minipage}{0.44\textwidth}
        \centering
        \begin{tabular}{ccc}
            \toprule
            AIA Filter~(\AA) & \(\Upsilon_L\) & \(\Upsilon_H\) \\
            \midrule
            0094 & 1.00 & 0.40 \\
            0131 & 0.60 & 0.25 \\
            0171 & 0.60 & 0.40 \\
            0193 & 0.70 & 0.40 \\
            0211 & 0.70 & 0.35 \\
            0304 & 0.90 & 0.50 \\
            0335 & 0.85 & 0.40 \\
            1600 & 0.80 & 0.40 \\
            1700 & 0.80 & 0.40 \\
            \bottomrule
        \end{tabular}
    \end{minipage}
    \caption{The \(\Upsilon\) redistribution family. Left: curves for different \(\Upsilon\) values, with a representative example \(\Upsilon=0.35\) in dark blue, alternatives in light blue, and red dashed lines showing standard gamma corrections for reference. Right: recommended default \(\Upsilon_L\) and \(\Upsilon_H\) values for AIA filters. Symmetric and asymmetric cases are supported.}
    \label{fig:upsilon_redistribution}
\end{figure}

\subsection{Upsilon \((\Upsilon)\) Redistribution} \label{sec:upsilon}

While RHEF provides strong perceptual enhancement, some datasets may still exhibit overly dominant bright or dark regions---especially when preparing images for display or outreach. \changed{To address this, we introduce an optional post-processing step, which maps the flat histogram to an arbitrary shape. We introduce the Upsilon \(\Upsilon\) redistribution function as a good choice, pulling the extreme values back towards the mean. While optional, we recommend enabling $\Upsilon$ when preparing images for interpretation, display scaling, or public communication. It retains the parameter-free ethos of RHEF while allowing intuitive and simple fine-tuning of the image contrast.}

\begin{figure}[hb!]
    \centering
    \includegraphics[width=1\linewidth]{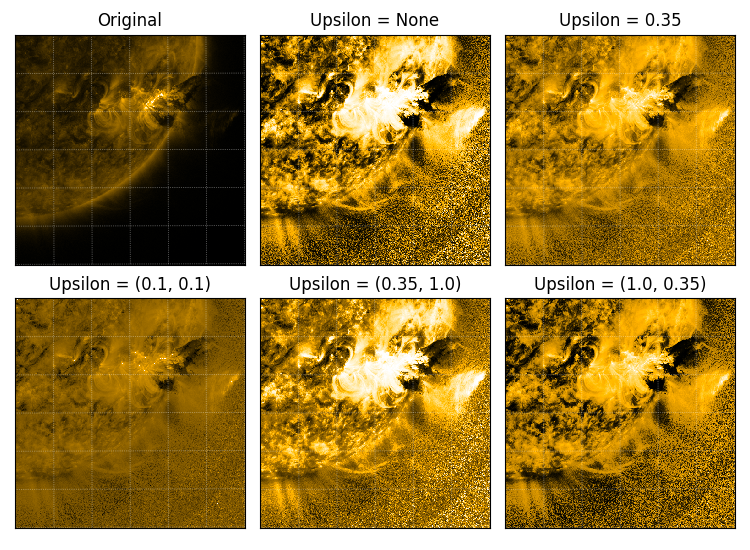}
    \caption{The sample 171 filter image from SunPy is displayed in the first panel. This is used to produce an RHEF filtered image, which is then treated with different values for Upsilon $\Upsilon$. \changed{Note that the top middle panel shows RHE without Upsilon.} Default values are given in Figure \ref{fig:upsilon_redistribution} for AIA channels.}
    \label{fig:upsilon}
\end{figure}

Mathematically, \(\Upsilon\) is defined as:
\begin{equation} \label{eq:sunback:upsilon}
I_{\mathrm{out}} = \Upsilon(I_{\mathrm{in}}, \Upsilon_L, \Upsilon_H) = \dfrac{1}{2}\,
\begin{cases}
\big(2 I_{\mathrm{in}}\big)^{\Upsilon_L}, & I_{\mathrm{in}} \le 0.5,\\[6pt]
\changedeq{2 - \big(2(1 - I_{\mathrm{in}})\big)^{\Upsilon_H}}, & I_{\mathrm{in}} > 0.5.
\end{cases}
\end{equation}
where \( I_{\mathrm{in}} \) and \( I_{\mathrm{out}} \) are the input and output intensities (both normalized to a range between 0 and 1) and \(\Upsilon_L\), \(\Upsilon_H\) independently control the redistribution curvature below and above the median, respectively. Figure~\ref{fig:upsilon_redistribution} compares \(\Upsilon\) curves (blue) to traditional gamma corrections (red), illustrating how \(\Upsilon\) preserves mid-range contrast while flattening both extremes. When \(\Upsilon = 1\), the transformation is linear; smaller values apply stronger compression, concentrating intensity values around 0.5. The effect of these choices can be seen in Figure \ref{fig:upsilon}.

Conceptually, \(\Upsilon\) reshapes the image histogram at each radius according to a smooth, tunable profile. In this work, we adopt this flexible double-sided gamma-correction-like curve that compresses the dynamic range around the median while preserving structural fidelity. Figure~\ref{fig:familyphoto}(d) shows the result of applying both RHEF and \(\Upsilon\).

This \(\Upsilon\) resembles a gamma correction in form, but with a crucial difference: it applies a double-sided transformation, mapping a standard gamma curve to the lower half of the normalized domain and an inverted gamma curve to the upper half. Unlike traditional gamma correction, which adjusts contrast in a single direction relative to a midpoint, \(\Upsilon\) performs symmetric compression about the median. This preserves midtone detail, controls both tails of the distribution, and offers finer control over perceptual brightness---reducing bias introduced by skewed histograms.

Whereas standard histogram equalization allocates equal population to each intensity bin, \(\Upsilon\) selectively compresses high and low tails while preserving contrast in perceptually important midtones. This avoids artificial extrema and improves tonal balance in both scientific and outreach applications. The function is scale-independent, computationally efficient, and compatible with any normalized dataset. Although designed to complement RHEF, it can be applied independently.

Default parameters (Figure~\ref{fig:upsilon_redistribution}) are empirically tuned for SDO/AIA and perform well across EUV channels. The rightmost column of Figure~\ref{fig:hist_compare_193} demonstrates \(\Upsilon\)'s impact when applied to an equalized image. Typical parameter ranges of \([0.5, 1.0]\) for \(\Upsilon_L\) and \([0.3, 0.5]\) for \(\Upsilon_H\) yield subjectively pleasing visual results for most EUV channels. The ability to tune these separately allows for asymmetric dynamic range compression when necessary.

The $\Upsilon$ redistribution is implemented in the \texttt{sunkit\_image.radial} module and may be applied to any normalized solar image. Full documentation and usage examples are available online\footnote{\url{https://docs.SunPy.org/projects/sunkit-image/en/stable/api/sunkit_image.radial.rhef.html}}. The redistribution function can also be used independently of RHEF by disabling filtering with \texttt{method=None}: This instructs the function to bypass histogram equalization and apply only the $\Upsilon$ transformation to the input map.

\section{Investigations and Results}
\label{sec:results}

\subsection{Performance Compared to Existing Methods}

This section compares the computational and visual performance of the nonlinear image enhancement techniques introduced earlier.
% Section~\ref{sec:results:histcompare} analyzes radial intensity distributions, while Section~\ref{sec:results:imgcompare} examines qualitative differences across key solar features.
Additional examples from instruments such as LASCO (onboard \textsc{SOHO}), SUVI (onboard \textsc{GOES}-R), K-Cor (operated by the Mauna Loa Solar Observatory), and synthetic data from \textsc{PUNCH} are provided in Appendix~\ref{sec:appendix_instruments}.

% \subsubsection{Radial Histograms}
% \label{sec:results:histcompare}

\begin{landscape}

\begin{figure}[ht!]
    \centering
    \includegraphics[width=\linewidth]{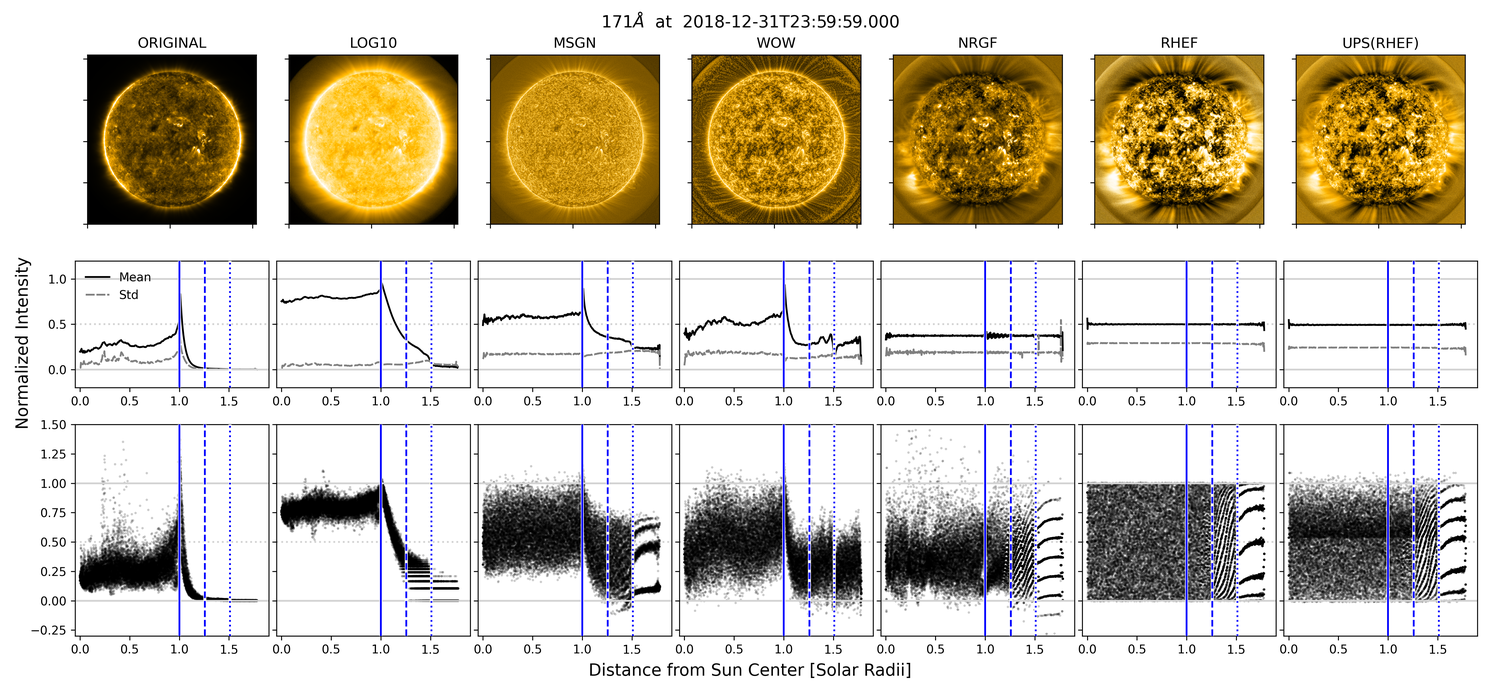}
    \caption{
        Comparison of enhancement techniques applied to a single AIA 171~\AA{} image.
        Top row: processed images. Middle row: radial profiles showing the mean and standard deviation of intensity versus distance from disk center.
        Bottom row: scatter plots showing the full distribution of normalized pixel intensities at each radius.
        % From left to right: NRGF, MSGN, RHEF, and RHEF + \(\Upsilon\).
        RHEF provides near-uniform radial statistics, while \(\Upsilon\) selectively compresses both tails of the intensity distribution.
        The vertical blue lines indicate the limb, detector edge, and optical edge, respectively.
        }
    \label{fig:hist_compare_193}
\end{figure}
\end{landscape}

Figure~\ref{fig:hist_compare_193} compares the outputs of several filtering techniques applied to a single AIA 171~\AA{} image. The top row shows the processed images, the middle row plots the radial mean and standard deviation of intensity, and the bottom row displays the normalized intensity distributions at each radial height using scatter plots.

The first \changed{two} columns provide baseline comparisons. The ``ORIGINAL'' frame corresponds to the raw data with exposure correction and instrument calibration applied.
The ``Log10'' column uses a logarithmic transform to compress the dynamic range. These serve as reference views for subsequent enhancements.

The \changed{third} column shows the result of MSGN, a multiscale enhancement technique that is \changed{insensitive} to radial geometry. MSGN reveals fine-scale structures across a broad range of spatial frequencies and naturally produces a balanced histogram, despite not explicitly addressing radial trends.

The fourth column shows WOW, the Wavelet Optimized Whitening filter. This filter equalizes the power across spatial scales, bringing out the fine structure in the image.

The fifth column presents the output of NRGF, which normalizes each pixel by \changed{subtracting the radial mean intensity and dividing by the standard deviation of the} intensity at its radius. This flattens the radial brightness trend off-limb well, but note that the algorithm was fundamentally not designed for use on-disk since there is no preferred direction at that height (See Section \ref{sec:mitigation} for more discussion). We include it here as a comparison because it often produces a similar looking result to RHEF but the radial histograms are completely different.

The sixth and seventh columns show the RHEF, with and without the optional \(\Upsilon\) redistribution step. RHEF remaps pixel intensities within each annulus based on percentile rank, producing a nearly uniform output profile that preserves ordinal structure and equalizes contrast without curve fitting. The radial statistics and scatter plots confirm this, with both mean and spread remaining consistent across radii. The final column adds the \(\Upsilon\) transformation, which smooths contrast by compressing the tails of the intensity distribution, aligning the output more closely with perceptual brightness ranges. As discussed in Section~\ref{sec:upsilon}, \(\Upsilon\) enables intuitive contrast adjustment without requiring detailed understanding of the tuning parameters.

% \remove{\subsubsection{Image Features}}

\begin{landscape}
    \begin{figure}
        \centering
        \includegraphics[width=1.0\linewidth]{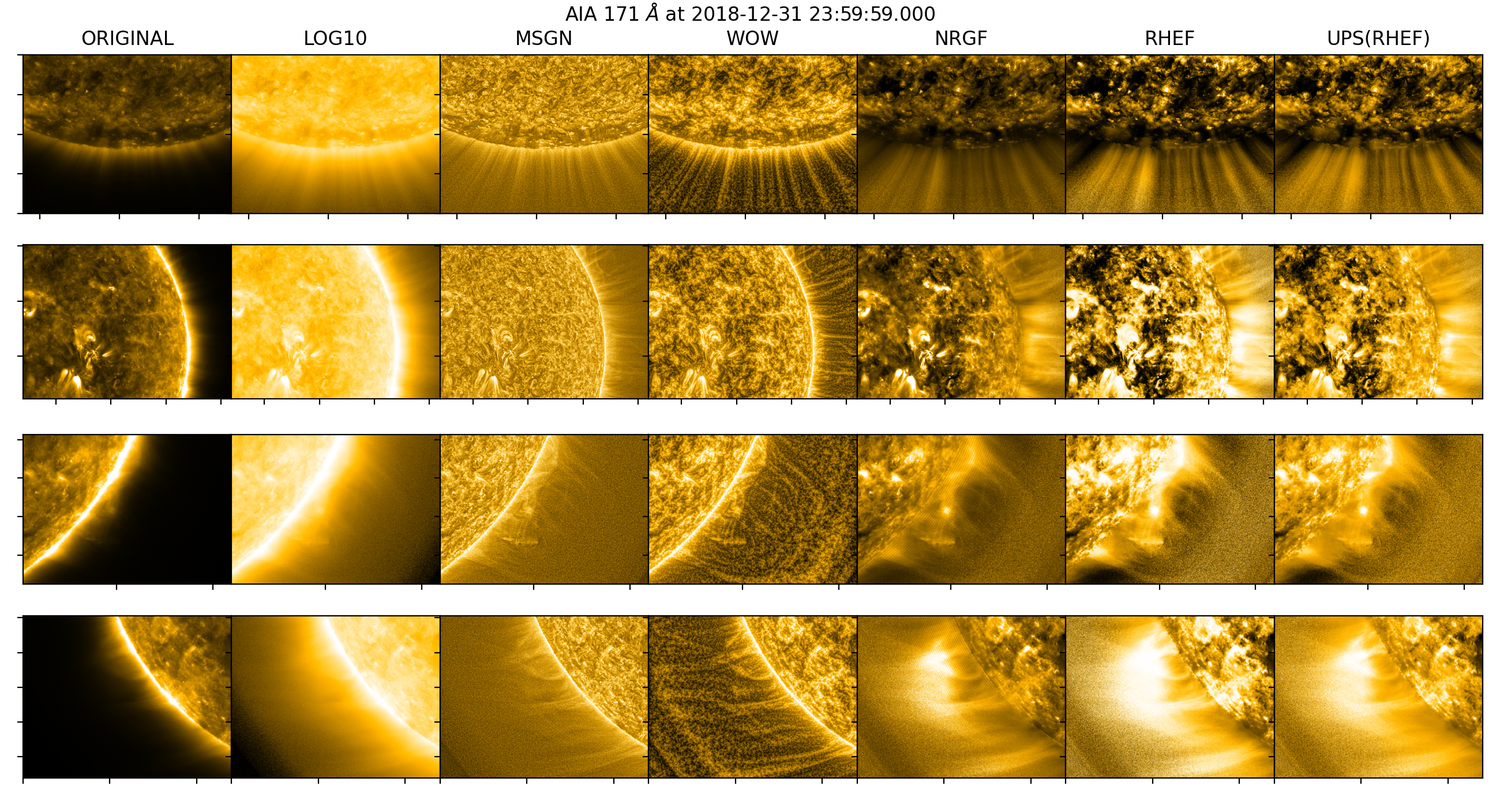}
        \caption{Comparison of enhancement techniques across diverse solar environments. \changed{Columns represent different filtering methods.} Each row shows a different feature class: polar plumes, equatorial corona and disk, a coronal loop, and a pseudo-streamer.}
        \label{fig:composite_comparison}
    \end{figure}
\end{landscape}

Figure~\ref{fig:composite_comparison} compares the performance of different filtering techniques across four representative solar contexts: polar plumes, the equatorial corona and disk, a coronal loop, and a pseudo-streamer. Each row corresponds to a distinct field of view, while each column shows the output of a specific nonlinear enhancement method. The figure highlights that each method has unique strengths, underscoring the importance of selecting an approach based on both the features of interest and how the algorithms affect them.

% MSGN enhances fine-scale features, particularly in textures like plume boundaries and coronal loops. However, as a high-pass filter, MSGN tends to suppress large-scale brightness gradients. In contrast, RHEF preserves global context while enhancing contrast through percentile remapping. When used together---either by applying RHEF to MSGN outputs or by linearly combining both---images exhibit both high-frequency detail and large-scale clarity. The final column illustrates such a blended approach.

\begin{figure}[ht!]
    \centering
    \includegraphics[width=0.495\linewidth]{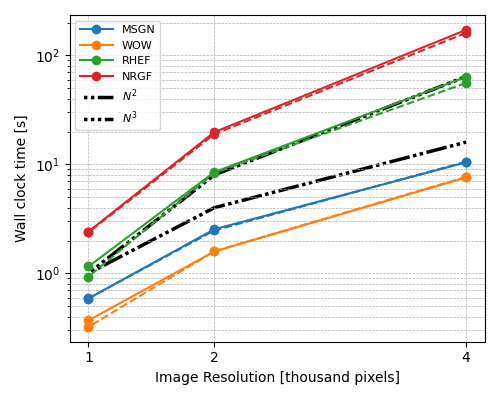}
    \includegraphics[width=0.495\linewidth]{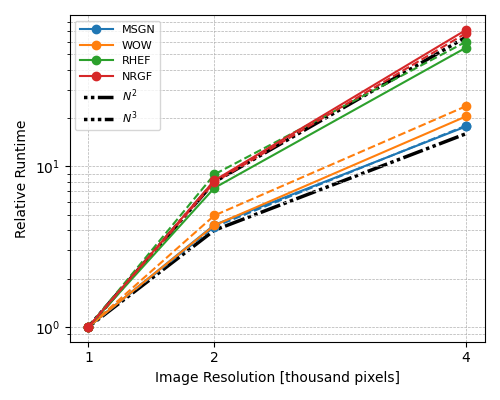} \\
    \includegraphics[width=0.8\linewidth]{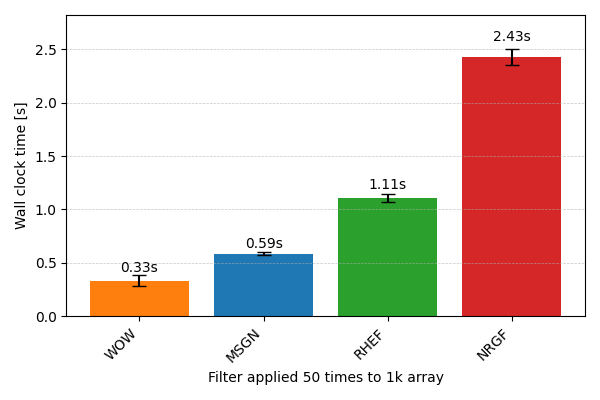}
    \caption{Runtime and scaling performance of selected filters. Lower values indicate faster execution per iteration. Top: Wall-clock time as a function of image resolution, plotted on log scales for absolute time (left) and \changed{normalized to show computational complexity} (right). Solid lines show upscaling of a small image; dashed lines show binning of a large image. Both rescaling steps were performed outside the timed loop, yielding similar scaling behavior. Decomposition-based methods scale as $N^2$ (black dash-double-dot line); RGF-like methods scale as $N^3$ (black dash-triple-dot line). Bottom: Mean wall-clock time and standard deviation across 50 runs on a $1k \times 1k$ image, displayed as a bar chart with error bars.}
    \label{fig:runtime_summary}
\end{figure}

% \subsubsection{Computational Complexity}
Figure~\ref{fig:runtime_summary} shows that RHEF scales as $N^3$, yet is still faster than the NRGF at every resolution. A 1024×1024 AIA image processes in around a second, while a 4096×4096 image completes in about a minute. This is without any bespoke optimization. The method is trivially parallelizable, with each annulus able to be processed independently. It requires no background models or time history, making it suitable for archival, single-shot, and near-real-time use.

Both WOW and MSGN complete faster because their convolution-based operations benefit from highly optimized, parallel implementations and direct 2D array processing. Convolution kernels, wavelet transforms, and Gaussian filters are efficiently computed using FFTs or separable operations, with excellent cache locality and vectorization. In contrast, radial methods like RHEF and NRGF incur additional overhead from mapping each pixel to radial coordinates, computing per-annulus statistics, and remapping back to the image plane---introducing non-local memory access patterns and extra interpolation steps. These differences make radial methods inherently less compatible with fast, general-purpose numerical libraries, despite their conceptual simplicity.

\subsection{Overall Evaluation of Radial Histogram Equalization}\label{sec:overall}

\subsubsection{Strengths}

Radial Histogram Equalization (RHEF) provides two principal benefits for solar image enhancement. First, it serves as a radial-gradient flattener, equalizing brightness across both the solar disk and limb to reveal structure in a single frame. Unlike background-subtraction techniques that rely on multi-frame context or long-term statistical models, RHEF operates in true single-shot mode. It can be applied immediately---even to the first frame in a series---which makes it especially valuable for real-time applications where a temporally-symmetric background model isn't available.

Second, RHEF compresses dynamic range by remapping intensity distributions within concentric annuli. Each annulus is independently ranked and normalized to the interval \([0, 1]\), producing a visually balanced output without the need for parameter tuning. This percentile-based approach naturally suppresses outliers while preserving ordinal structure. For instance, an input array such as \([1, 13, 500000, 600]\) might be remapped to ranks \([1, 2, 4, 3]\), flattening brightness disparities without losing relational contrast.

Each radial zone's median maps to a fixed output value---typically 0.5---which helps ensure frame-to-frame consistency in animations and synoptic series. This improves temporal coherence and minimizes flicker. The algorithm is also computationally lightweight: a 1024×1024 image typically processes in \changed{around} one second, and even full-resolution 4096×4096 data can be filtered in \changed{about a minute}. Because each annulus is processed independently, the algorithm is trivially parallelizable and well-suited for both real-time visualization and large-scale batch processing.

\subsubsection{Optimal Applications}

RHEF is suitable for analyses that don't rely on calibrated intensity, as its nonlinear, spatially variable percentile remapping disrupts the physical meaning of pixel values. This precludes its use in tasks like emission measure reconstruction, temperature diagnostics, or density modeling.

However, RHEF excels in morphology-focused workflows, where structure, shape, and texture matter more than absolute radiance. It is well suited for feature-driven studies such as plume analysis \citep{Deforest2001}, coronal hole detection \citep{Krista2009}, coronal-loop tracing \citep{Aschwanden2010}, and wave tracking \citep{Weberg2018}.

By compressing dynamic range, RHEF enhances the visibility of fine structures critical to such analyses. That said, because RHEF does not address or remove any constant background features, \remove{low SNR features should be removed with }a background subtraction method \changed{should be applied before RHEF is used if the desired signal is much less intense than the measured signal.} Appendix~\ref{sec:appendix_lasco} demonstrates this phenomenon using LASCO data.

Users should note that percentile remapping can subtly alter the contrast field, especially near intensity gradients---potentially affecting threshold-based methods where fixed cutoffs are applied. The optional \(\Upsilon\) redistribution function (Section~\ref{sec:upsilon}) further reshapes brightness levels, emphasizing midtones and compressing distribution tails. When used, its effects should be validated within the specific context of downstream detection or classification tasks.

\subsubsection{Challenges and Potential Mitigation Strategies}\label{sec:mitigation}

\changed{RHEF, like other radial filters, can exhibit binning artifacts. These are most often seen as faint on-disk rings when a narrow annulus is dominated by a bright or dark feature. Percentile ranking is more robust than mean-based RGFs, but extreme cases can still bias an annulus and produce temporal jitter as features drift between bins. Widening annuli reduces this problem but at the cost of reintroducing the global radial gradient as banding.}

% A practical improvement is to compute percentile statistics over a broader window while applying the normalization within a narrower sliding annulus, similar to a radial Savitzky--Golay filter \citep{Savitzky1964}.}

\begin{landscape}
    \begin{figure}
        \centering
        \includegraphics[width=1.0\linewidth]{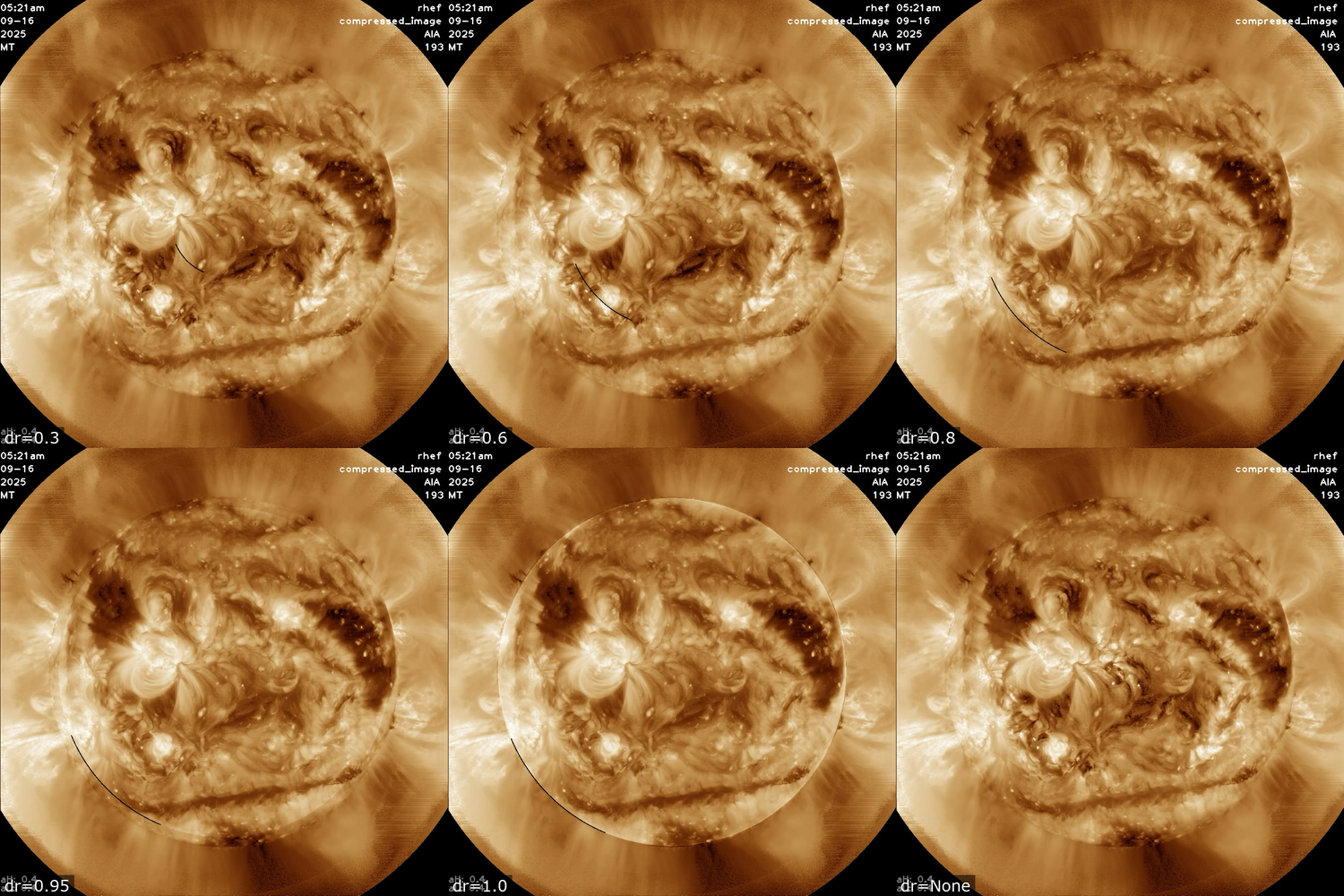}
        \caption{One approach to mitigate small-number statistics near disk center is to treat all heights below the specified $dr$ as a single bin. Circle segments at $dr = [0.3, 0.6, 0.8, 0.95, 1.0, None]$ are overplotted in each of the six panels respectively. The bottom row shows clear ring artifacts at $r=dr$, while the top row appears mostly artifact-free. A video version of this Figure, available in the online journal, highlights that seams also appear in the top row to varying degrees, with $dr=0.6$ providing the most effective balance.}
        \label{fig:disk_study}
    \end{figure}
\end{landscape}

\changed{Artifacts are strongest near disk center, where few pixels per annulus yield noisy statistics. A simple mitigation is to treat the innermost region (e.g., \(r<0.3\,R_{\odot}\)) as a single normalization zone. We implemented this option and tested several inner-disk radii \(dr\). Figure~\ref{fig:disk_study} compares six trial values from 0 to \(1\,R_\odot\), each annotated in the lower-left, with an arc at \(r=dr\) to mark the seam. Very large radii (\(dr=0.95\) and \(1.0\,R_\odot\)) produce limb-adjacent rings relative to the nominal RHE (bottom right). Intermediate values appear similar in stills, while the accompanying video reveals subtle seams.}

In animations, full-FOV RHE produces faint flickering rings near disk center as structures cross narrow annular boundaries. The disk-reduced variant suppresses this jitter but introduces a stationary seam at the chosen transition radius \(r=dr\).
Physically, the solar limb is the only self-consistent boundary, but it is also the most visually distracting.
Thus, users must trade a central flicker for a seam. For AIA 171, we find \(dr\approx0.6\,R_\odot\) offers the best balance, minimizing visibility while maintaining temporal stability.
Because no clear rationale exists for choosing a specific radius, this parameter is not implemented in SunPy, though users can replicate it by passing \texttt{radial\_bin\_edges} with the first values as, e.g., \([0, 0.6, 0.61, ...]\). Another solution is selecting an \texttt{application\_radius} $\geq 1.0\,R_\odot$, which will black out the entire image below that height.

\section{Conclusion}
\label{sec:conclusion}

\subsection{Summary}

After reviewing solar image filtering families in Section~\ref{sec:background}, we introduced the Radial Histogram Equalizing Filter (RHEF; Section~\ref{sec:rhef}), a hybrid, parameter-free method for enhancing high dynamic range (HDR) solar imagery, along with the optional Upsilon redistribution function (\(\Upsilon\), Section~\ref{sec:upsilon}), a symmetrized extension of standard gamma correction. Together, these tools offer a fast, intuitive approach for visualizing faint coronal structures across both the solar disk and extended limb, requiring minimal user tuning.

RHEF applies percentile-based remapping within concentric annuli, compressing dynamic range while preserving ordinal relationships, and works consistently across spatial scales and geometries. As shown in Section~\ref{sec:results}, it compares favorably to established techniques, producing smooth, normalized outputs that excel for morphology-focused tasks such as segmentation, feature tracking, and visual inspection, even though it is unsuitable for photometric applications. By making faint structures---such as plumes, streamers, and waves---more accessible, RHEF supports both scientific interpretation and operational forecasting.

Both RHEF and \(\Upsilon\) are implemented in the open-source \texttt{sunkit\_image.radial} Python package\footnote{Source code: \url{https://github.com/SunPy/sunkit-image}; Gallery: \url{https://docs.SunPy.org/projects/sunkit-image/en/stable/generated/gallery/index.html}} \citep{Barnes2020}. Additional reproducible workflows, documentation, and instrument-specific examples are provided online and in Appendix~\ref{sec:appendix_instruments}, and contributions and feedback are welcome via the project's GitHub repository. We have also provided an ``as-is'' IDL implementation that was graciously produced by the anonymous reviewer \citep{dr_gilly_2025_17226415}.

\subsection{Future Work}

Planned enhancements include adding CLAHE-style contrast limiting \citep{Pisano1998} and smoothing percentile transitions across neighboring annuli \changed{with the goal of improving} performance in low-SNR regions and reducing artifacts near the disk center and limb. We are also exploring hybrid combinations of RHEF with multiscale filters such as MSGN to better preserve fine-scale texture alongside global gradient suppression, and evaluating RHEF-enhanced imagery as a baseline for magnetic field extrapolation comparisons (e.g., PFSS and HMI-based models) to support studies of magnetic topology and wave propagation. Looking ahead, RHEF-enhanced imagery may also benefit machine learning applications in solar feature detection, classification, or event forecasting.

We are interested in integrating RHEF into resources like JHelioviewer and IDL SolarSoft, broadening accessibility for mission operations, real-time exploration, and outreach across widely used platforms.

% As with many desirable improvements to the filter, that is outside of the scope of this work for the time being.

Off-limb, particularly in low signal-to-noise environments like coronal holes or the extended streamer belt, noise can become exaggerated by the filter. Ranking algorithms treat all pixel differences equally, regardless of their physical significance, so faint structures and background noise are enhanced with equal weight. Pre-filtering with denoising techniques---such as the noise-gating method of \citet{DeForest2017}---can reduce this amplification. Of course integrating several frames in time will also increase the SNR and improve the noise-performance of the image. Future implementations may explore CLAHE-style contrast limiting \citep{Zuiderveld1994} or histogram smoothing \citep{Pizer1987, Yadav2014} to mitigate such effects more robustly.

\begin{acks}
    This work was supported by the National Aeronautics and Space Administration (NASA) under grants NNX15AW33G, NNX16AG87G, and 80NSSC20K1319, and by the National Science Foundation (NSF) under grants 1540094 and 1613207.
    The project was also supported by the Southwest Research Institute, providing overhead funding that enables proposal-driven research.
    The authors have no competing interests to declare that are relevant to the content of this article.
    All data used in this research is available from the Joint Science Operations Center (JSOC) at Stanford University, and can be accessed in a variety of modalities, from programmatic to web-based.
    Automated AI grammar and consistency tools were used during final copyediting; all analyses, figures, and scientific claims are the authors' own.
    We thank Nabil Freij for reviewing and assisting with the integration of our tools into SunPy \citep{Barnes2020}. Thanks also to the anonymous reviewer for their many improvements to this article, and for providing the IDL implementation of the core RHEF algorithm.
\end{acks}

\vfill
% \clearpage
\appendix

\section{Instrument-Specific Demonstrations}
\label{sec:appendix_instruments}

To demonstrate the versatility of the Radial Histogram Equalizing Filter (RHEF), we present example applications across a diverse set of solar and heliospheric instruments. These include both space-based and ground-based observatories, spanning ultraviolet, white-light, and synthetic imagery. In each case, the same core algorithm is applied using default parameters unless otherwise noted. These examples illustrate how RHEF generalizes across wavelengths, spatial resolutions, and instrumental architectures---underscoring its utility as a universal enhancement technique.

\clearpage
\subsection{SDO/AIA Color Composites}
\label{sec:results:rgb}

\begin{figure}[hb!]
    \centering
    \includegraphics[width=0.8\linewidth]{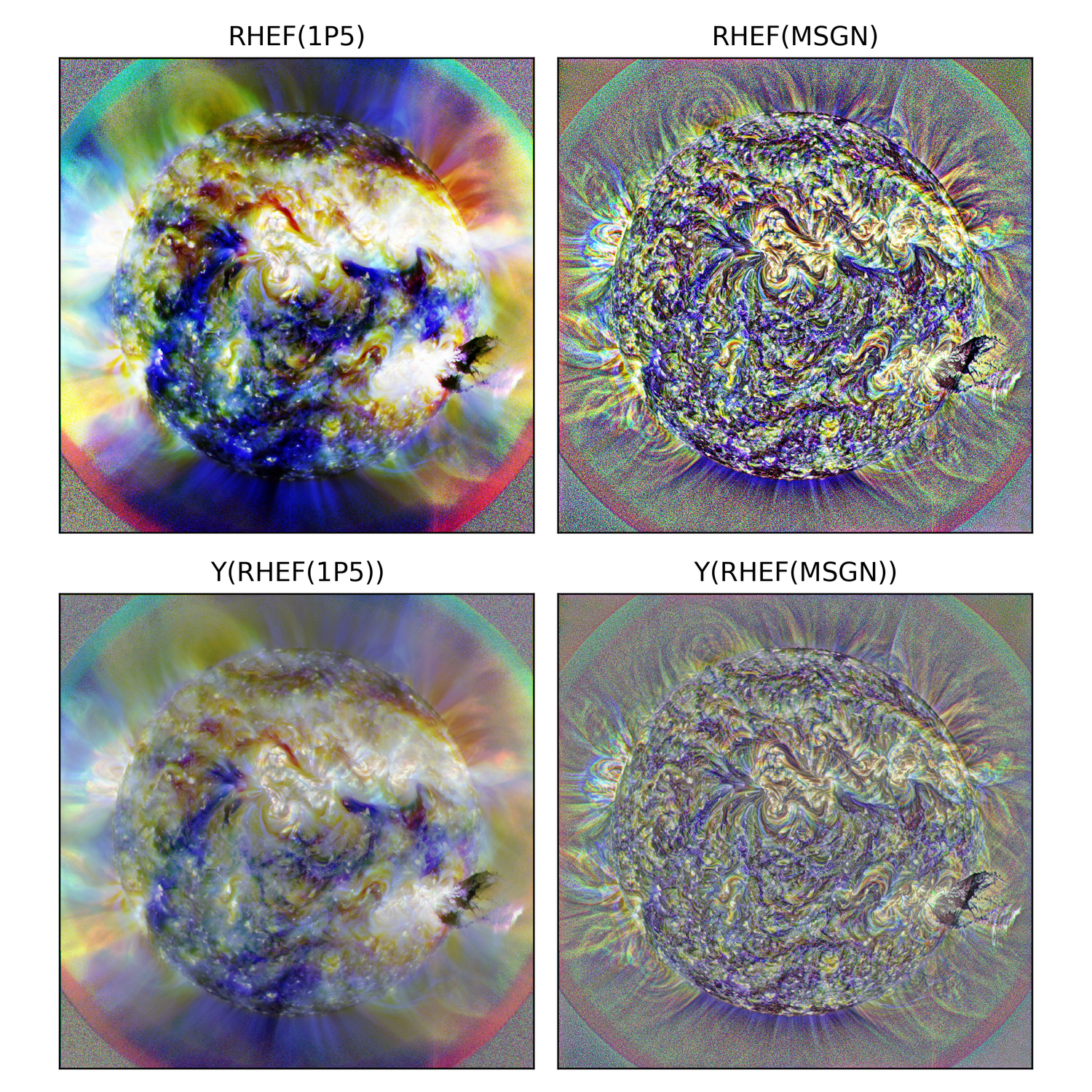}
    \caption{RGB composites using AIA 171~\AA{} (blue), 193~\AA{} (green), and 211~\AA{} (red), with each channel filtered independently. Left: RHEF applied to Level 1.5 data. Right: RHEF applied to MSGN-enhanced inputs. Top row: uncompressed; bottom row: with \(\Upsilon\) redistribution.}
    \label{fig:composite}
\end{figure}

Figure~\ref{fig:composite} shows RGB composites constructed from three SDO/AIA EUV channels: 171~\AA{} (blue), 193~\AA{} (green), and 211~\AA{} (red). Each color channel is independently enhanced using RHEF and then composited.

In the top row, RHEF is applied directly to Level 1.5 data. While this yields vivid and structurally rich renderings, brightness from dominant channels can oversaturate the composite. The bottom row includes an additional \(\Upsilon\) redistribution step applied to each channel, which compresses intensity extremes and improves tonal balance---especially in faint peripheral regions.

The second column in each row illustrates an alternative pipeline where RHEF is applied to images previously enhanced with Multiscale Gaussian Normalization (MSGN). This hybrid approach combines global gradient suppression from RHEF with fine-scale texture enhancement from MSGN, yielding outputs suited for both scientific analysis and public outreach.

While $\Upsilon$ redistribution compresses overly bright or dark regions, it also enhances midtone contrast, which can be helpful for morphological analysis. For scientific use, the use of $\Upsilon$ aids structural interpretation; for outreach, uncompressed versions may preserve the dramatic contrast often favored in visualizations.

\clearpage
\subsection{GOES-16/SUVI}
\label{sec:appendix_suvi}

The Solar Ultraviolet Imager (SUVI) onboard the GOES-R series provides full-disk EUV observations from geostationary orbit \citep{Martinez-Galarce2013, Seaton2018a, Tadikonda2019b}. Though its spatial resolution and point-spread function are coarser than instruments like AIA, SUVI excels at monitoring large-scale structures across a wide field of view.

\begin{figure}[hb!]
    \centering
    \includegraphics[width=0.8\linewidth]{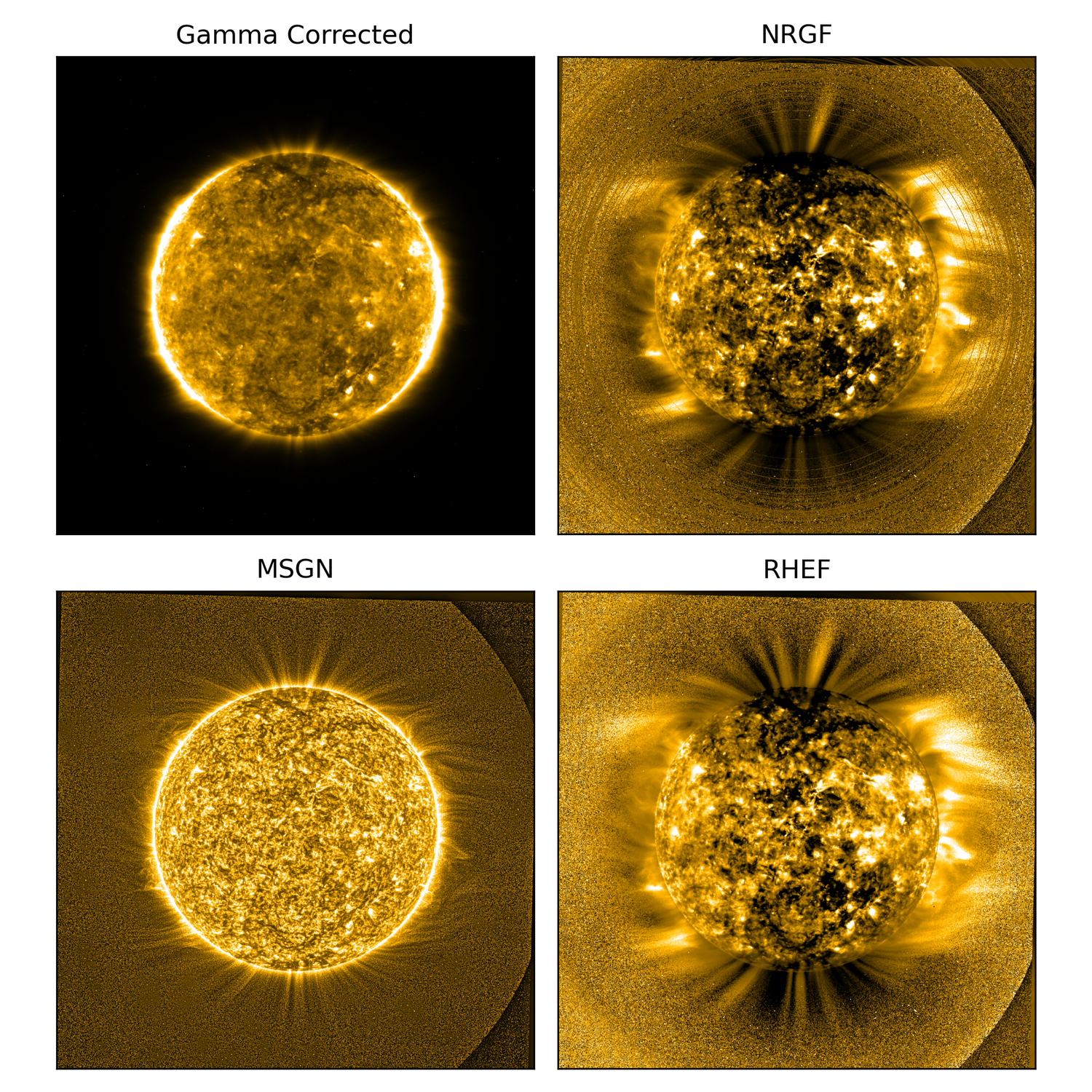}
    \caption{Comparison of NOAA SUVI 171~\AA{} image enhancement methods. All panels are clipped at the 4th and 99th percentiles. RHEF reveals faint off-limb structure while preserving detail without additional scaling. Timestamp: 2020-07-10}
    \label{fig:suvi}
\end{figure}

Figure~\ref{fig:suvi} compares gamma correction, NRGF, MSGN, and RHEF enhancements applied to the same SUVI 171~\AA{} image. All panels are clipped at the 4th and 99th percentiles for consistent visual scaling. RHEF differs from the others in that it performs contrast normalization intrinsically, using local percentile ranks without requiring global post-scaling.

MSGN reveals fine-scale structure, while NRGF suppresses radial gradients but suffers from low contrast. RHEF achieves balance between the two, enhancing both on-disk and off-limb features without parameter tuning. This demonstrates that RHEF remains effective even in instruments with lower signal-to-noise or broader spatial response.

\clearpage
\subsection{MLSO/K-Cor}
\label{sec:appendix_kcor}

The COronal Solar Magnetism Observatory (COSMO) K-Coronagraph (K-Cor) at Mauna Loa Solar Observatory captures high-cadence, white-light observations of the low solar corona. These images are essential for studying streamers, CME onset, and dynamic plasma flows near the limb \citep{Burkepile2013,Tomczyk2022}.

\begin{figure}[hb!]
    \centering
    \includegraphics[width=0.8\linewidth]{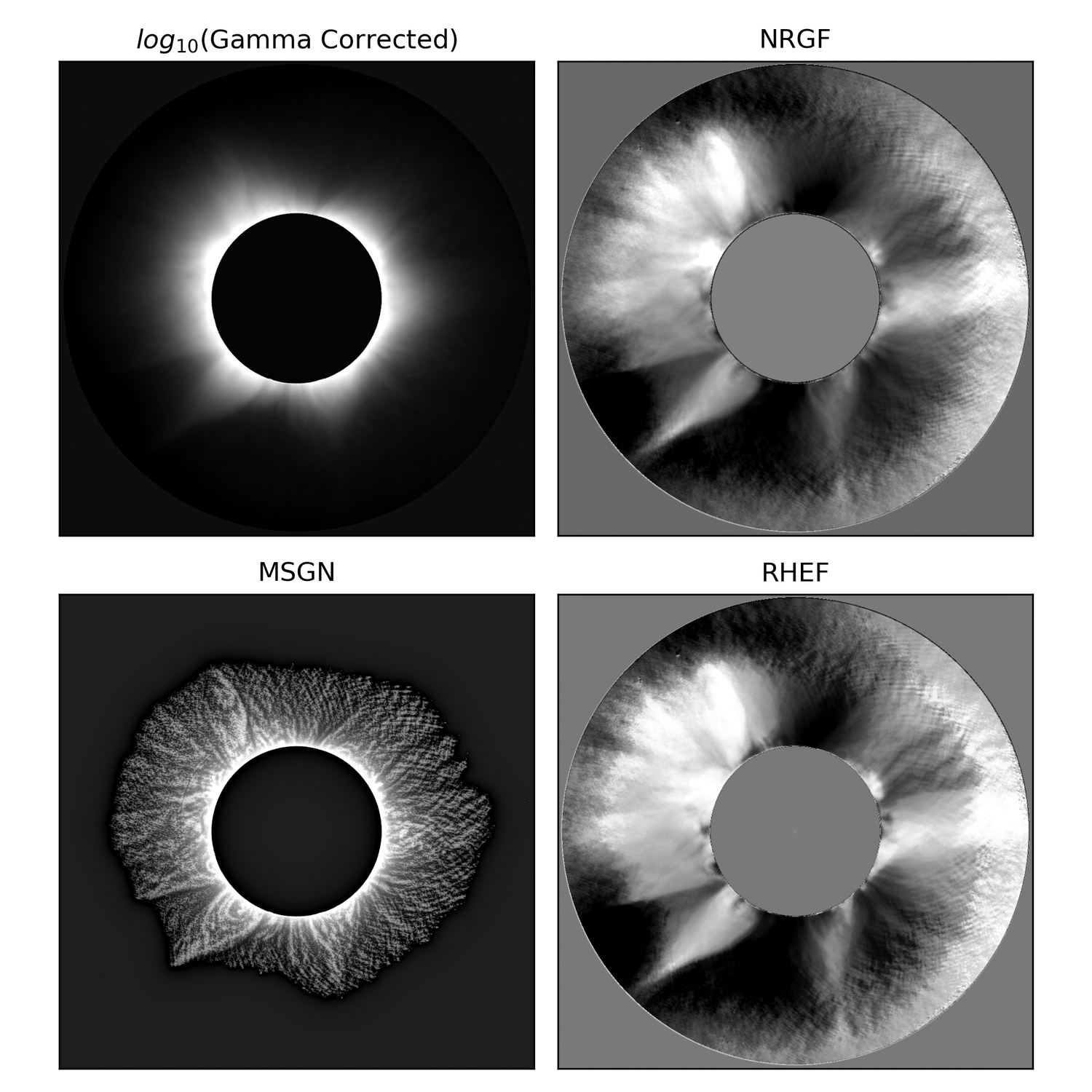}
    \caption{Application of RHEF to a K-Cor white-light observation taken on 2022-11-25 at 12:51:31. Streamer features and off-limb structures become more prominent without requiring background subtraction. }
    \label{fig:kcor-rhef}
\end{figure}

Figure~\ref{fig:kcor-rhef} shows a K-Cor frame enhanced using RHEF. Although the instrument observes broadband visible light rather than EUV, RHEF's percentile-based normalization is \changed{independent of} wavelength. This allows the method to effectively enhance faint, spatially extended structures such as streamer stalks and transient outflows without requiring a temporal background model or tuning.
RHEF proves especially useful in this regime, where high dynamic range and low SNR make traditional filters less effective.

\clearpage
\subsection{SOHO/LASCO C3}
\label{sec:appendix_lasco}

\begin{figure}[hb!]
  \centering
    \includegraphics[width=0.49\linewidth]{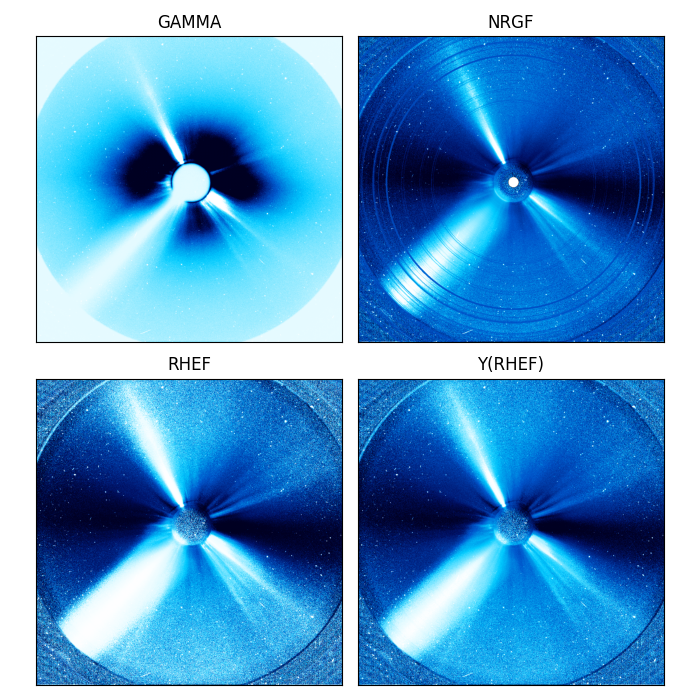}
    \includegraphics[width=0.49\linewidth]{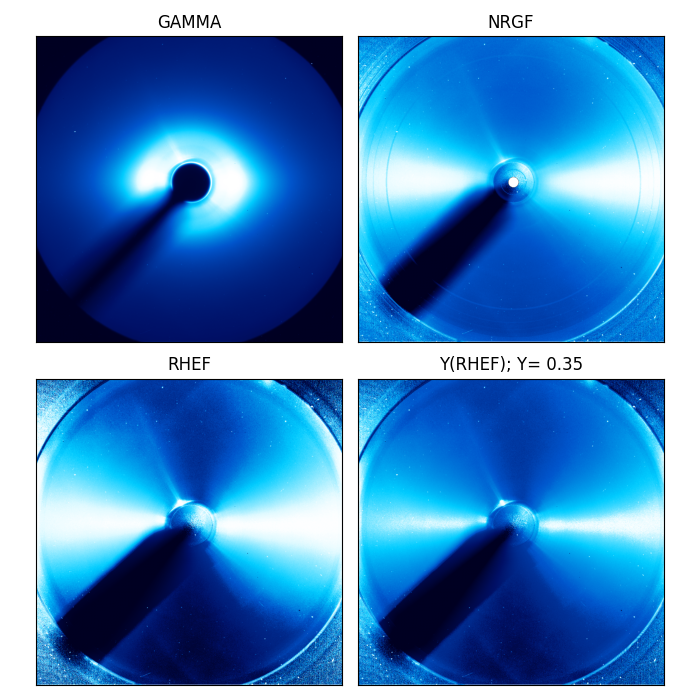}
  \caption{\changed{LASCO C3 white-light imagery from 2022/04/22 at 00:06:05, enhanced using different methods. The left set of four shows the background-subtracted image, while the right set shows images without background subtraction.}}
  \label{fig:lasco}
\end{figure}

The LASCO C3 coronagraph aboard SOHO observes the corona from 3.7-30~\(R_\odot\), enabling studies of CMEs, streamers, and large-scale coronal dynamics \citep{Brueckner1995}. Background subtraction follows the running background method of \citet{Morrill2006}. Figure~\ref{fig:lasco} compares an unprocessed Level 0 image (gamma-corrected) with outputs from NRGF, and RHEF and \(\Upsilon(RHEF)\). The right panels show filtering without background subtraction; the left panels include it. The significantly \changed{better-visualized coronal features on the left} highlight the importance of understanding both the data and the appropriate filtering approach.

RHEF reveals streamer substructure and faint outflows beyond 5~\(R_\odot\), features largely obscured in raw or gamma-corrected images. Unlike NRGF, which applies a global min/max scale, RHEF adjusts contrast dynamically per annulus, enhancing tonal balance and structural detail without requiring post-normalization.

\clearpage
\subsection{Synthetic PUNCH Data}
\label{sec:appendix_punch}

The Polarimeter to UNify the Corona and Heliosphere (PUNCH) is a recently-launched space mission designed to provide high cadence wide-field imagery of the Sun's corona \cite{Deforest2022a}.

\begin{figure}[hb!]
    \centering
    \includegraphics[width=0.8\linewidth]{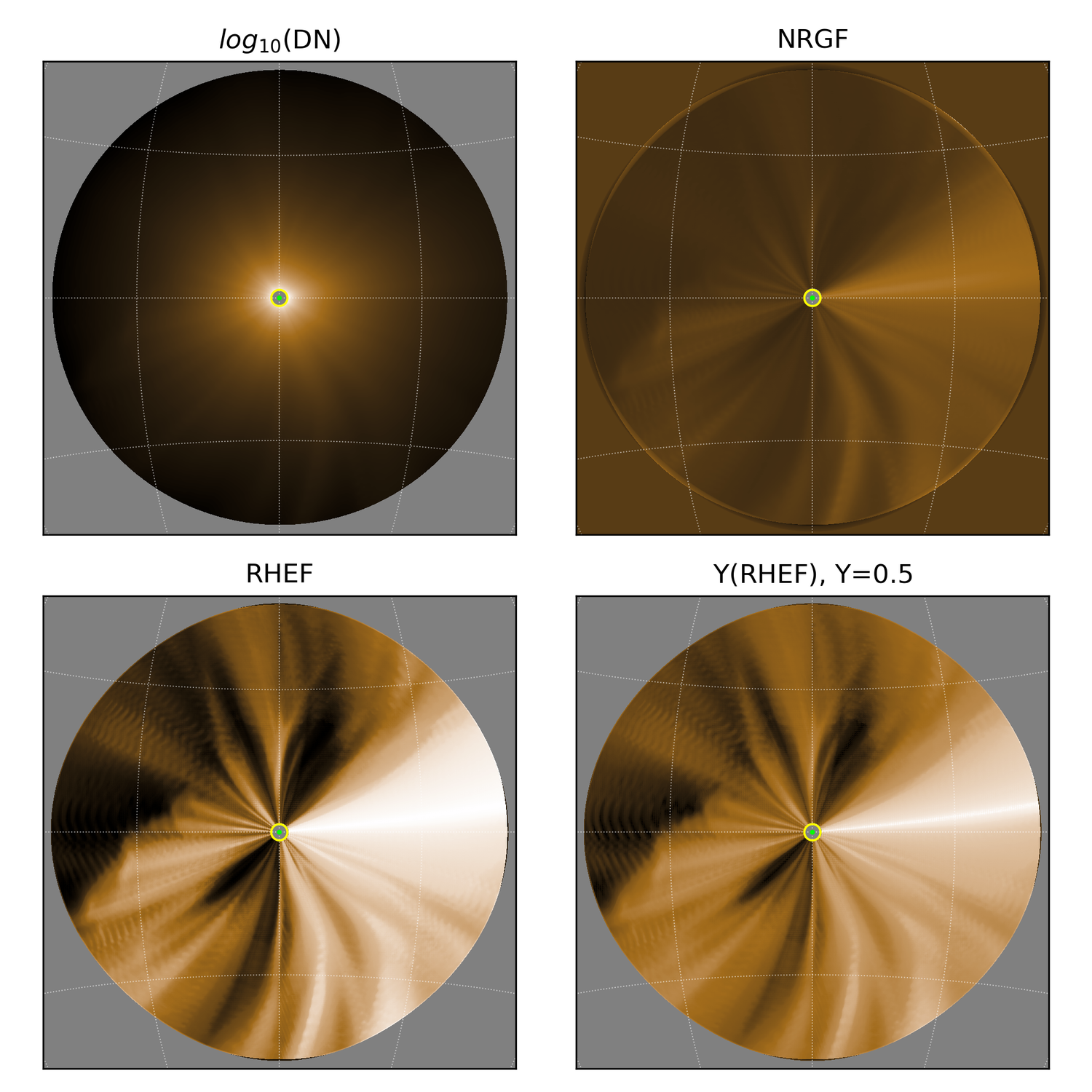}
    \caption{Synthetic white-light solar wind imagery processed with (1) logarithmic scaling, (2) NRGF, (3) RHEF, and (4) RHEF + \(\Upsilon\). RHEF preserves streamer boundaries and small-scale density gradients, while \(\Upsilon\) enhances tonal smoothness.}
    \label{fig:PUNCH}
\end{figure}

Figure~\ref{fig:PUNCH} shows RHEF applied to synthetic wide-field white-light images produced by the FORWARD \citep{Gibson2016b} radiative transfer code, illuminating solar wind structures from the GAMERA MHD model \citep{Mostafavi2022,Zhang2019a} as presented by \citet{Provornikova2024}. These simulations replicate heliospheric Thomson-scattered observations of features such as streamers and density inhomogeneities.

The baseline panel uses logarithmic scaling; subsequent panels apply NRGF, RHEF, and RHEF with \(\Upsilon\) redistribution. \remove{Compared to NRGF, which globally flattens gradients and downplays fine details, RHEF better preserves boundaries and faint outflows. Adding \(\Upsilon\) further improves tonal coherence without compromising structure.} These results highlight RHEF's potential for real-time feature detection in wide-angle missions like NASA's PUNCH, where single-frame enhancement will aid solar wind and CME tracking.

\clearpage
\section{AIA Synoptic Dataset}
\label{sec:appendix_synoptic}

For most examples in this work, we used full-resolution AIA Level 1.5 data. However, the Joint Science Operations Center (JSOC) also offers a downsampled AIA synoptic series---a lightweight resource ideal for prototyping and rapid testing.
This dataset includes reduced-resolution (1024x1024) EUV images at a 2-minute cadence, derived from 4096x4096 Level 1.5 data by averaging 4x4 blocks of valid pixels. The resulting pixel values are in DN per 0.6 arcsec pixel. Although these files are not served via the DRMS interface, we have implemented a \texttt{FIDOClient} for SunPy's FIDO API (v7.0 and up) to streamline access.
Because of their small size and consistent cadence, synoptic images are excellent for iterating image-processing code at scale. An example query is shown in Listing~\ref{code:aiasynopticclient}. These synoptic directories can also be browsed directly at \url{https://jsoc1.stanford.edu/data/aia/synoptic/}.

\begin{lstlisting}[language=Python, caption={Accessing the AIA Synoptic Dataset using FIDO}, label={code:aiasynopticclient}]
from astropy import units as u
from sunpy.net import Fido, attrs as a

search_result = Fido.search(
    a.Time("2015-06-06", "2015-06-06"),
    a.Level("1.5s"),  #  <--- This is the new line
    a.Sample(1 * u.hour),
    a.Wavelength(171 * u.angstrom),
    a.Instrument("AIA")
)
file_path = Fido.fetch(search_result)
\end{lstlisting}

% \clearpage

\section{Python Usage Example}\label{sec:appendix_python}
A minimal usage example is shown below in Listing \ref{code:rhef}, and more example scripts\footnote{\url{https://docs.SunPy.org/projects/sunkit-image/en/stable/generated/gallery/index.html}} and usage patterns are provided in the online documentation\citep{Gilly2024}\footnote{\url{https://docs.SunPy.org/projects/sunkit-image/en/stable/api/sunkit_image.radial.rhef.html}}.

\begin{lstlisting}[language=Python, caption={Using the RHEF on a sample SunPy map with Upsilon disabled}, label={code:rhef}]
from sunpy.map import Map
from sunpy.data.sample import AIA_171_IMAGE
from sunkit_image.radial import rhef
import matplotlib.pyplot as plt

aia_map = Map(AIA_171_IMAGE)
filtered_map = rhef(aia_map, upsilon=None)

filtered_map.plot()
plt.show()
\end{lstlisting}

% \section{IDL Implementation}
% \label{sec:appendix_IDL}

% We thank the reviewer for providing a version of RHEF that is written in IDL, presented here in Listing \ref{code:IDL_RHEF}. We hope to implement this code into SolarSoft in the future for ease of use.

% \begin{lstlisting}[language=IDL, caption={IDL Implementation of RHEF}, label={code:IDL_RHEF}]
% pro rhef, f
%     read_sdo,f,hdr,im
%     t=systime(1)
%     x=findgen(hdr.naxis1)-hdr.crpix1
%     y=findgen(hdr.naxis2)-hdr.crpix2
%     xx=rebin(x,hdr.naxis1,hdr.naxis2)
%     yy=rebin(transpose(y),hdr.naxis1,hdr.naxis2)
%     r=sqrt(xx^2+yy^2)
%     h=histogram(r,binsize=1,rev=revind,loc=rh)
%     nh=n_elements(h)
%     imout=im*0.
%     for index_hist=0,nh-1 do begin
%         if h[index_hist] eq 0 then continue
%         res=revind[revind[index_hist]:revind[index_hist+1]-1]
%         n=n_elements(res)
%         imnow=im[res]
%         imout[res]=interpol(findgen(n)/(n-1),imnow[sort(imnow)],imnow)
%     endfor
%     yl=0.7
%     yh=0.4
%     indl=where(imout le 0.5,comp=indh)
%     imout[indl]=0.5*((2*imout[indl])^yl)
%     imout[indh]=0.5*(2-(2*(1-imout[indh]))^yh)
%     print,'Time elapsed: ',systime(1)-t
%     tvscl,rebin(imout,1024,1024)
% end

% ; Usage Example
% IDL> rhef, 'aia.lev1_euv_12s_2024-01-01T00_00_00Z_193.fits', YL=0.6, YH=0.45

% \end{lstlisting}

\clearpage

\bibliographystyle{spr-mp-sola}  % Replace with the appropriate style for solarphysics
\bibliography{library}  % List your .bib files here, without the extension

\begin{thebibliography}{81}
% BibTex style file: spr-mp-sola.bst, v2.05, 2022-11-16
\ifx\bisbn     \undefined \def\bisbn  #1{ISBN #1}\fi
\ifx\binits    \undefined \def\binits#1{#1}\fi
\ifx\bauthor   \undefined \def\bauthor#1{#1}\fi
\ifx\batitle   \undefined \def\batitle#1{#1}\fi
\ifx\bjtitle   \undefined \def\bjtitle#1{\textit{#1}}\fi
\ifx\bvolume   \undefined \def\bvolume#1{\textbf{#1}}\fi
\ifx\byear     \undefined \def\byear#1{#1}\fi
\ifx\bissue    \undefined \def\bissue#1{#1}\fi
\ifx\bfpage    \undefined \def\bfpage#1{#1}\fi
\ifx\blpage    \undefined \def\blpage #1{#1}\fi
\ifx\burl      \undefined \def\burl#1{#1}\fi
\ifx\href      \undefined \def\href#1#2{#2}\fi
\ifx\betal     \undefined \def\betal{et al.}\fi
\ifx\bctitle   \undefined \def\bctitle#1{#1}\fi
\ifx\beditor   \undefined \def\beditor#1{#1}\fi
\ifx\bbtitle   \undefined \def\bbtitle#1{\textit{#1}}\fi
\ifx\bedition  \undefined \def\bedition#1{#1}\fi
\ifx\bseriesno \undefined \def\bseriesno#1{\textbf{#1}}\fi
\ifx\blocation \undefined \def\blocation#1{#1}\fi
\ifx\bsertitle \undefined \def\bsertitle#1{\textit{#1}}\fi
\ifx\bsnm      \undefined \def\bsnm#1{#1}\fi
\ifx\bsuffix   \undefined \def\bsuffix#1{#1}\fi
\ifx\bparticle \undefined \def\bparticle#1{#1}\fi
\ifx\barticle  \undefined \def\barticle#1{}\fi
\ifx\binstitute  \undefined \def\binstitute#1{#1}\fi
\ifx\bpublisher  \undefined \def\bpublisher#1{#1}\fi
\ifx\doiurl    \undefined \def\doiurl#1{\href{#1}{DOI}}\fi
\makeatletter
\def\safeHref#1#2#3{\in@{http}{#2}\ifin@\href{#2}{#3}\else\href{#1#2}{#3}\fi}
\makeatother
\ifx\adsurl    \undefined
  \def\adsurl#1{\safeHref{https://ui.adsabs.harvard.edu/abs/}{#1}{ADS}}\fi
\ifx\arxivurl  \undefined
  \def\arxivurl#1{\safeHref{http://arxiv.org/abs/}{#1}{arXiv}}\fi
\ifx\botherref \undefined \def\botherref#1{}\fi
\ifx\url       \undefined \def\url#1{#1}\fi
\ifx\bchapter  \undefined \def\bchapter#1{}\fi
\ifx\bbook     \undefined \def\bbook#1{}\fi
\ifx\bcomment  \undefined \def\bcomment#1{#1}\fi
\ifx\oauthor   \undefined \def\oauthor#1{#1}\fi
\ifx\citeauthoryear \undefined\def \citeauthoryear#1{#1}\fi
\def\endbibitem {}
\ifx\bconflocation  \undefined \def\bconflocation#1{#1} \fi

\bibitem[\protect\citeauthoryear{Alzate and Morgan}{2017}]{Alzate2017}
\begin{barticle}
\bauthor{\bsnm{Alzate}, \binits{N.}},
\bauthor{\bsnm{Morgan}, \binits{H.}}:
\byear{2017},
\batitle{{Identification of Low Coronal Sources of “Stealth” Coronal Mass
  Ejections Using New Image Processing Techniques}}.
\bjtitle{Astrophys. J.}
\bvolume{840},
\bfpage{103}.
\doiurl{https://doi.org/10.3847/1538-4357/aa6caa}.
\end{barticle}
\endbibitem

\bibitem[\protect\citeauthoryear{Andiani, Indrajit, and
  Dazki}{2024}]{Andiani2024}
\begin{barticle}
\bauthor{\bsnm{Andiani}},
\bauthor{\bsnm{Indrajit}, \binits{R.E.}},
\bauthor{\bsnm{Dazki}, \binits{E.}}:
\byear{2024},
\batitle{{Multi-Scale Adaptive Contrast Enhancement (MSACE) for Color Images: A
  Comparative Analysis with Conventional Techniques}}.
\bjtitle{J. Inf. Syst. Eng. Manag.}
\bvolume{10},
\bfpage{49}.
\doiurl{https://doi.org/10.52783/jisem.v10i8s.956}.
\burl{https://jisem-journal.com/index.php/journal/article/view/956}.
\end{barticle}
\endbibitem

\bibitem[\protect\citeauthoryear{Aschwanden}{2010}]{Aschwanden2010}
\begin{barticle}
\bauthor{\bsnm{Aschwanden}, \binits{M.J.}}:
\byear{2010},
\batitle{{A code for automated tracing of coronal loops approaching visual
  perception}}.
\bjtitle{Sol. Phys.}
\bvolume{262},
\bfpage{399}.
\bisbn{1120701095316}.
\doiurl{https://doi.org/10.1007/s11207-010-9531-6}.
\burl{https://ui.adsabs.harvard.edu/abs/2010SoPh..262..399A/abstract
  https://link.springer.com/article/10.1007/s11207-010-9531-6}.
\end{barticle}
\endbibitem

\bibitem[\protect\citeauthoryear{Auch{\`{e}}re et~al.}{2023a}]{Auchere2023b}
\begin{barticle}
\bauthor{\bsnm{Auch{\`{e}}re}, \binits{F.}},
\bauthor{\bsnm{Berghmans}, \binits{D.}},
\bauthor{\bsnm{Dumesnil}, \binits{C.}},
\bauthor{\bsnm{Halain}, \binits{J.P.}},
\bauthor{\bsnm{Mercier}, \binits{R.}},
\bauthor{\bsnm{Rochus}, \binits{P.}},
\bauthor{\bsnm{Delmotte}, \binits{F.}},
\bauthor{\bsnm{Fran{\c{c}}ois}, \binits{S.}},
\bauthor{\bsnm{Hermans}, \binits{A.}},
\bauthor{\bsnm{Hervier}, \binits{V.}},
\bauthor{\bsnm{Kraaikamp}, \binits{E.}},
\bauthor{\bsnm{Meltchakov}, \binits{E.}},
\bauthor{\bsnm{Morinaud}, \binits{G.}},
\bauthor{\bsnm{Philippon}, \binits{A.}},
\bauthor{\bsnm{Smith}, \binits{P.J.}},
\bauthor{\bsnm{Stegen}, \binits{K.}},
\bauthor{\bsnm{Verbeeck}, \binits{C.}},
\bauthor{\bsnm{Zhang}, \binits{X.}},
\bauthor{\bsnm{Andretta}, \binits{V.}},
\bauthor{\bsnm{Abbo}, \binits{L.}},
\bauthor{\bsnm{Buchlin}, \binits{E.}},
\bauthor{\bsnm{Frassati}, \binits{F.}},
\bauthor{\bsnm{Gissot}, \binits{S.}},
\bauthor{\bsnm{Gyo}, \binits{M.}},
\bauthor{\bsnm{Harra}, \binits{L.}},
\bauthor{\bsnm{Jerse}, \binits{G.}},
\bauthor{\bsnm{Landini}, \binits{F.}},
\bauthor{\bsnm{Mierla}, \binits{M.}},
\bauthor{\bsnm{Nicula}, \binits{B.}},
\bauthor{\bsnm{Parenti}, \binits{S.}},
\bauthor{\bsnm{Renotte}, \binits{E.}},
\bauthor{\bsnm{Romoli}, \binits{M.}},
\bauthor{\bsnm{Russano}, \binits{G.}},
\bauthor{\bsnm{Sasso}, \binits{C.}},
\bauthor{\bsnm{Sch{\"{u}}hle}, \binits{U.}},
\bauthor{\bsnm{Schmutz}, \binits{W.}},
\bauthor{\bsnm{Soubri{\'{e}}}, \binits{E.}},
\bauthor{\bsnm{Susino}, \binits{R.}},
\bauthor{\bsnm{Teriaca}, \binits{L.}},
\bauthor{\bsnm{West}, \binits{M.}},
\bauthor{\bsnm{Zhukov}, \binits{A.N.}}:
\byear{2023}a,
\batitle{{Beyond the disk: EUV coronagraphic observations of the Extreme
  Ultraviolet Imager on board Solar Orbiter}}.
\bjtitle{Astron. Astrophys.}
\bvolume{674},
\bfpage{A127}.
\doiurl{https://doi.org/10.1051/0004-6361/202346039}.
\burl{https://ui.adsabs.harvard.edu/abs/2023A{\%}5C{\%}5C{\&}A...674A.127A/abstract}.
\end{barticle}
\endbibitem

\bibitem[\protect\citeauthoryear{Auch{\`{e}}re et~al.}{2023b}]{Auchere2023a}
\begin{barticle}
\bauthor{\bsnm{Auch{\`{e}}re}, \binits{F.}},
\bauthor{\bsnm{Soubri{\'{e}}}, \binits{E.}},
\bauthor{\bsnm{Pelouze}, \binits{G.}},
\bauthor{\bsnm{Buchlin}}:
\byear{2023}b,
\batitle{{Image enhancement with wavelet-optimized whitening}}.
\bjtitle{Astron. Astrophys.}
\bvolume{670}.
\doiurl{https://doi.org/10.1051/0004-6361/202245345}.
\end{barticle}
\endbibitem

\bibitem[\protect\citeauthoryear{Barnes et~al.}{2020}]{Barnes2020}
\begin{barticle}
\bauthor{\bsnm{Barnes}, \binits{W.T.}},
\bauthor{\bsnm{Bobra}, \binits{M.G.}},
\bauthor{\bsnm{Christe}, \binits{S.D.}},
\bauthor{\bsnm{Freij}, \binits{N.}},
\bauthor{\bsnm{Hayes}, \binits{L.A.}},
\bauthor{\bsnm{Ireland}, \binits{J.}},
\bauthor{\bsnm{Mumford}, \binits{S.}},
\bauthor{\bsnm{Perez-Suarez}, \binits{D.}},
\bauthor{\bsnm{Ryan}, \binits{D.F.}},
\bauthor{\bsnm{Shih}, \binits{A.Y.}},
\bauthor{\bsnm{Chanda}, \binits{P.}},
\bauthor{\bsnm{Glogowski}, \binits{K.}},
\bauthor{\bsnm{Hewett}, \binits{R.}},
\bauthor{\bsnm{Hughitt}, \binits{V.K.}},
\bauthor{\bsnm{Hill}, \binits{A.}},
\bauthor{\bsnm{Hiware}, \binits{K.}},
\bauthor{\bsnm{Inglis}, \binits{A.}},
\bauthor{\bsnm{Kirk}, \binits{M.S.F.}},
\bauthor{\bsnm{Konge}, \binits{S.}},
\bauthor{\bsnm{Mason}, \binits{J.P.}},
\bauthor{\bsnm{Maloney}, \binits{S.A.}},
\bauthor{\bsnm{Murray}, \binits{S.A.}},
\bauthor{\bsnm{Panda}, \binits{A.}},
\bauthor{\bsnm{Park}, \binits{J.}},
\bauthor{\bsnm{Pereira}, \binits{T.M.D.}},
\bauthor{\bsnm{Reardon}, \binits{K.}},
\bauthor{\bsnm{Savage}, \binits{S.}},
\bauthor{\bsnm{Sipőcz}, \binits{B.M.}},
\bauthor{\bsnm{Stansby}, \binits{D.}},
\bauthor{\bsnm{Jain}, \binits{Y.}},
\bauthor{\bsnm{Taylor}, \binits{G.}},
\bauthor{\bsnm{Yadav}, \binits{T.}},
\bauthor{\bsnm{Rajul}},
\bauthor{\bsnm{Dang}, \binits{T.K.}}:
\byear{2020},
\batitle{{The SunPy Project: Open Source Development and Status of the Version
  1.0 Core Package}}.
\bjtitle{Astrophys. J.}
\bvolume{890},
\bfpage{68}.
\doiurl{https://doi.org/10.3847/1538-4357/ab4f7a}.
\burl{https://ui.adsabs.harvard.edu/abs/2020ApJ...890...68S/abstract
  https://iopscience.iop.org/article/10.3847/1538-4357/ab4f7a/meta}.
\end{barticle}
\endbibitem

\bibitem[\protect\citeauthoryear{Boe et~al.}{2021}]{Boe2021}
\begin{barticle}
\bauthor{\bsnm{Boe}, \binits{B.}},
\bauthor{\bsnm{Habbal}, \binits{S.R.}},
\bauthor{\bsnm{Downs}, \binits{C.}},
\bauthor{\bsnm{Druckm{\"{u}}ller}, \binits{M.}}:
\byear{2021},
\batitle{{The Color and Brightness of the F-corona Inferred from the 2019 July
  2 Total Solar Eclipse}}.
\bjtitle{Astrophys. J.}
\bvolume{912},
\bfpage{44}.
\doiurl{https://doi.org/10.3847/1538-4357/abea79}.
\end{barticle}
\endbibitem

\bibitem[\protect\citeauthoryear{Boerner et~al.}{2012}]{Boerner2012a}
\begin{barticle}
\bauthor{\bsnm{Boerner}, \binits{P.}},
\bauthor{\bsnm{Edwards}, \binits{C.}},
\bauthor{\bsnm{Lemen}, \binits{J.R.}},
\bauthor{\bsnm{Rausch}, \binits{A.}},
\bauthor{\bsnm{Schrijver}, \binits{C.}},
\bauthor{\bsnm{Shine}, \binits{R.}},
\bauthor{\bsnm{Shing}, \binits{L.}},
\bauthor{\bsnm{Stern}, \binits{R.}},
\bauthor{\bsnm{Tarbell}, \binits{T.}},
\bauthor{\bsnm{Title}, \binits{A.}},
\bauthor{\bsnm{Wolfson}, \binits{C.J.}},
\bauthor{\bsnm{Soufli}, \binits{R.}},
\bauthor{\bsnm{Spiller}, \binits{E.}},
\bauthor{\bsnm{Gullikson}, \binits{E.}},
\bauthor{\bsnm{McKenzie}, \binits{D.}},
\bauthor{\bsnm{Windt}, \binits{D.}},
\bauthor{\bsnm{Golub}, \binits{L.}},
\bauthor{\bsnm{Podgorski}, \binits{W.}},
\bauthor{\bsnm{Testa}, \binits{P.}},
\bauthor{\bsnm{Weber}, \binits{M.}}:
\byear{2012},
\batitle{{Initial calibration of the Atmospheric Imaging Assembly (AIA) on the
  Solar Dynamics Observatory (SDO)}}.
\bjtitle{Sol. Phys.}
\bvolume{9781461436},
\bfpage{41}.
\bisbn{9781461436737}.
\doiurl{https://doi.org/10.1007/978-1-4614-3673-7_4}.
\burl{https://ui.adsabs.harvard.edu/abs/2012SoPh..275...41B/abstract
  http://download.springer.com/static/pdf/90/art{\%}253A10.1007{\%}252Fs11207-011-9804-8.pdf?originUrl=http{\%}3A{\%}2F{\%}2Flink.springer.com{\%}2Farticle{\%}2F10.1007{\%}2Fs11207-011-9804-8{\&}token2=exp=1493775003{~}acl=}.
\end{barticle}
\endbibitem

\bibitem[\protect\citeauthoryear{Brueckner et~al.}{1995}]{Brueckner1995}
\begin{barticle}
\bauthor{\bsnm{Brueckner}, \binits{G.E.}},
\bauthor{\bsnm{Howard}, \binits{R.A.}},
\bauthor{\bsnm{Koomen}, \binits{M.J.}},
\bauthor{\bsnm{Korendyke}, \binits{C.M.}},
\bauthor{\bsnm{Michels}, \binits{D.J.}},
\bauthor{\bsnm{Moses}, \binits{J.D.}},
\bauthor{\bsnm{Socker}, \binits{D.G.}},
\bauthor{\bsnm{Dere}, \binits{K.P.}},
\bauthor{\bsnm{Lamy}, \binits{P.L.}},
\bauthor{\bsnm{Llebaria}, \binits{A.}},
\bauthor{\bsnm{Bout}, \binits{M.V.}},
\bauthor{\bsnm{Schwenn}, \binits{R.}},
\bauthor{\bsnm{Simnett}, \binits{G.M.}},
\bauthor{\bsnm{Bedford}, \binits{D.K.}},
\bauthor{\bsnm{Eyles}, \binits{C.J.}}:
\byear{1995},
\batitle{{The Large Angle Spectroscopic Coronagraph (LASCO) - Visible light
  coronal imaging and spectroscopy}}.
\bjtitle{Sol. Phys.}
\bvolume{162},
\bfpage{357}.
\doiurl{https://doi.org/10.1007/BF00733434}.
\burl{https://ui.adsabs.harvard.edu/abs/1995SoPh..162..357B/abstract}.
\end{barticle}
\endbibitem

\bibitem[\protect\citeauthoryear{Burkepile et~al.}{2013}]{Burkepile2013}
\begin{barticle}
\bauthor{\bsnm{Burkepile}, \binits{J.}},
\bauthor{\bsnm{DeWijn}, \binits{A.}},
\bauthor{\bsnm{Tomczyk}, \binits{S.}},
\bauthor{\bsnm{Sewell}, \binits{S.}},
\bauthor{\bsnm{Gallagher}, \binits{D.}},
\bauthor{\bsnm{Sutherland}, \binits{L.}},
\bauthor{\bsnm{Card}, \binits{G.}},
\bauthor{\bsnm{Lecinski}, \binits{A.}},
\bauthor{\bsnm{Larson}, \binits{B.}},
\bauthor{\bsnm{Nelson}, \binits{P.}},
\bauthor{\bsnm{Huang}, \binits{P.}},
\bauthor{\bsnm{Kolinski}, \binits{D.}},
\bauthor{\bsnm{Sitongia}, \binits{L.}},
\bauthor{\bsnm{Stueben}, \binits{A.}},
\bauthor{\bsnm{Berkey}, \binits{B.}},
\bauthor{\bsnm{Burkepile}, \binits{J.}},
\bauthor{\bsnm{DeWijn}, \binits{A.}},
\bauthor{\bsnm{Tomczyk}, \binits{S.}},
\bauthor{\bsnm{Sewell}, \binits{S.}},
\bauthor{\bsnm{Gallagher}, \binits{D.}},
\bauthor{\bsnm{Sutherland}, \binits{L.}},
\bauthor{\bsnm{Card}, \binits{G.}},
\bauthor{\bsnm{Lecinski}, \binits{A.}},
\bauthor{\bsnm{Larson}, \binits{B.}},
\bauthor{\bsnm{Nelson}, \binits{P.}},
\bauthor{\bsnm{Huang}, \binits{P.}},
\bauthor{\bsnm{Kolinski}, \binits{D.}},
\bauthor{\bsnm{Sitongia}, \binits{L.}},
\bauthor{\bsnm{Stueben}, \binits{A.}},
\bauthor{\bsnm{Berkey}, \binits{B.}}:
\byear{2013},
\batitle{{The COSMO K-Coronagraph}}.
\bjtitle{EGUGA}
\bvolume{15},
\bfpage{EGU2013}.
\burl{https://ui.adsabs.harvard.edu/abs/2013EGUGA..1511810B/abstract}.
\end{barticle}
\endbibitem

\bibitem[\protect\citeauthoryear{Byrne et~al.}{2012}]{Byrne2012}
\begin{barticle}
\bauthor{\bsnm{Byrne}, \binits{J.P.}},
\bauthor{\bsnm{Morgan}, \binits{H.}},
\bauthor{\bsnm{Habbal}, \binits{S.R.}},
\bauthor{\bsnm{Gallagher}, \binits{P.T.}}:
\byear{2012},
\batitle{{Automatic detection and tracking of coronal mass ejections. II.
  Multiscale filtering of coronagraph images}}.
\bjtitle{Astrophys. J.}
\bvolume{752},
\bfpage{145}.
\doiurl{https://doi.org/10.1088/0004-637X/752/2/145}.
\end{barticle}
\endbibitem

\bibitem[\protect\citeauthoryear{Byrne et~al.}{2014}]{Byrne2014}
\begin{barticle}
\bauthor{\bsnm{Byrne}, \binits{J.P.}},
\bauthor{\bsnm{Morgan}, \binits{H.}},
\bauthor{\bsnm{Seaton}, \binits{D.B.}},
\bauthor{\bsnm{Bain}, \binits{H.M.}},
\bauthor{\bsnm{Habbal}, \binits{S.R.}}:
\byear{2014},
\batitle{{Bridging EUV and white-light observations to inspect the initiation
  phase of a “two-stage” solar eruptive event}}.
\bjtitle{Sol. Phys.}
\bvolume{289},
\bfpage{4545}.
\doiurl{https://doi.org/10.1007/s11207-014-0585-8}.
\end{barticle}
\endbibitem

\bibitem[\protect\citeauthoryear{Cho et~al.}{2019}]{Cho2019}
\begin{barticle}
\bauthor{\bsnm{Cho}, \binits{K.-S.}},
\bauthor{\bsnm{Cho}, \binits{I.-H.}},
\bauthor{\bsnm{Nakariakov}, \binits{V.M.}},
\bauthor{\bsnm{Yurchyshyn}, \binits{V.B.}},
\bauthor{\bsnm{Yang}, \binits{H.}},
\bauthor{\bsnm{Kim}, \binits{Y.-H.}},
\bauthor{\bsnm{Kumar}, \binits{P.}},
\bauthor{\bsnm{Magara}, \binits{T.}}:
\byear{2019},
\batitle{{Oscillation of a Small H $\alpha$ Surge in a Solar Polar Coronal
  Hole}}.
\bjtitle{Astrophys. J.}
\bvolume{877},
\bfpage{L1}.
\doiurl{https://doi.org/10.3847/2041-8213/ab1eb5}.
\end{barticle}
\endbibitem

\bibitem[\protect\citeauthoryear{Cranmer}{2020}]{Cranmer2020c}
\begin{barticle}
\bauthor{\bsnm{Cranmer}, \binits{S.R.}}:
\byear{2020},
\batitle{{Heating Rates for Protons and Electrons in Polar Coronal Holes:
  Empirical Constraints from the Ultraviolet Coronagraph Spectrometer}}.
\bjtitle{Astrophys. J.}
\bvolume{900},
\bfpage{105}.
\doiurl{https://doi.org/10.3847/1538-4357/abab04}.
\burl{https://ui.adsabs.harvard.edu/abs/2020ApJ...900..105C/abstract}.
\end{barticle}
\endbibitem

\bibitem[\protect\citeauthoryear{DeForest}{2017}]{DeForest2017}
\begin{barticle}
\bauthor{\bsnm{DeForest}, \binits{C.E.}}:
\byear{2017},
\batitle{{Noise-gating to Clean Astrophysical Image Data}}.
\bjtitle{Astrophys. J.}
\bvolume{838},
\bfpage{155}.
\doiurl{https://doi.org/10.3847/1538-4357/aa67f1}.
\burl{http://dx.doi.org/10.3847/1538-4357/aa67f1
  https://ui.adsabs.harvard.edu/abs/2017ApJ...838..155D/abstract}.
\end{barticle}
\endbibitem

\bibitem[\protect\citeauthoryear{DeForest, Howard, and
  McComas}{2014}]{DeForest2014}
\begin{barticle}
\bauthor{\bsnm{DeForest}, \binits{C.E.}},
\bauthor{\bsnm{Howard}, \binits{T.A.}},
\bauthor{\bsnm{McComas}, \binits{D.J.}}:
\byear{2014},
\batitle{{Inbound waves in the solar corona: A direct indicator of Alfv{\'{e}}n
  surface location}}.
\bjtitle{Astrophys. J.}
\bvolume{787},
\bfpage{124}.
\doiurl{https://doi.org/10.1088/0004-637X/787/2/124}.
\burl{http://arxiv.org/abs/1404.3235
  http://dx.doi.org/10.1088/0004-637X/787/2/124
  https://iopscience.iop.org/article/10.1088/0004-637X/787/2/124
  https://iopscience.iop.org/article/10.1088/0004-637X/787/2/124/meta}.
\end{barticle}
\endbibitem

\bibitem[\protect\citeauthoryear{DeForest, Lamy, and
  Llebaria}{2001}]{Deforest2001}
\begin{barticle}
\bauthor{\bsnm{DeForest}, \binits{C.E.}},
\bauthor{\bsnm{Lamy}, \binits{P.L.}},
\bauthor{\bsnm{Llebaria}, \binits{A.}}:
\byear{2001},
\batitle{{Solar Polar Plume Lifetime and Coronal Hole Expansion: Determination
  from Long‐Term Observations}}.
\bjtitle{Astrophys. J.}
\bvolume{560},
\bfpage{490}.
\doiurl{https://doi.org/10.1086/322497}.
\burl{http://iopscience.iop.org/article/10.1086/322497/pdf
  https://ui.adsabs.harvard.edu/abs/2001ApJ...560..490D/abstract}.
\end{barticle}
\endbibitem

\bibitem[\protect\citeauthoryear{Deforest et~al.}{2022}]{Deforest2022a}
\begin{bchapter}
\bauthor{\bsnm{Deforest}, \binits{C.E.}},
\bauthor{\bsnm{Killough}, \binits{R.}},
\bauthor{\bsnm{Gibson}, \binits{S.}},
\bauthor{\bsnm{Henry}, \binits{A.}},
\bauthor{\bsnm{Case}, \binits{T.}},
\bauthor{\bsnm{Beasley}, \binits{M.}},
\bauthor{\bsnm{Laurent}, \binits{G.}},
\bauthor{\bsnm{Colaninno}, \binits{R.}},
\bauthor{\bsnm{Waltham}, \binits{N.}}:
\byear{2022},
\bctitle{{Polarimeter to UNify the Corona and Heliosphere (PUNCH): Science,
  Status, and Path to Flight}}.
In: \bbtitle{IEEE Aerosp. Conf. Proc.}
\bseriesno{2022-March},
\bpublisher{Institute of Electrical and Electronics Engineers (IEEE)},
\bfpage{1,1}.
\bisbn{9781665437608}.
\doiurl{https://doi.org/10.1109/AERO53065.2022.9843340}.
\burl{https://ui.adsabs.harvard.edu/abs/2022aero.confE...1D/abstract}.
\end{bchapter}
\endbibitem

\bibitem[\protect\citeauthoryear{{Del Zanna} and Mason}{2018}]{DelZanna2018}
\begin{botherref}
\oauthor{\bsnm{{Del Zanna}}, \binits{G.}},
\oauthor{\bsnm{Mason}, \binits{H.E.}}:
2018,
\textit{{Solar UV and X-ray spectral diagnostics}},
Springer.
\doiurl{https://doi.org/10.1007/s41116-018-0015-3}.
\url{https://doi.org/10.1007/s41116-018-0015-3
  https://ui.adsabs.harvard.edu/abs/2018LRSP...15....5D/abstract}.
\end{botherref}
\endbibitem

\bibitem[\protect\citeauthoryear{Druckm{\"{u}}ller}{2013}]{Druckmuller2013}
\begin{barticle}
\bauthor{\bsnm{Druckm{\"{u}}ller}, \binits{M.}}:
\byear{2013},
\batitle{{A noise adaptive fuzzy equalization method for processing solar
  extreme ultraviolet images}}.
\bjtitle{Astrophys. Journal, Suppl. Ser.}
\bvolume{207},
\bfpage{25}.
\doiurl{https://doi.org/10.1088/0067-0049/207/2/25}.
\burl{https://ui.adsabs.harvard.edu/abs/2013ApJS..207...25D/abstract}.
\end{barticle}
\endbibitem

\bibitem[\protect\citeauthoryear{Druckm{\"{u}}ller, Habbal, and
  Morgan}{2014}]{Druckmuller2014}
\begin{barticle}
\bauthor{\bsnm{Druckm{\"{u}}ller}, \binits{M.}},
\bauthor{\bsnm{Habbal}, \binits{S.R.}},
\bauthor{\bsnm{Morgan}, \binits{H.}}:
\byear{2014},
\batitle{{Discovery of a new class of coronal structures in white light eclipse
  images}}.
\bjtitle{Astrophys. J.}
\bvolume{785},
\bfpage{14}.
\doiurl{https://doi.org/10.1088/0004-637X/785/1/14}.
\burl{https://ui.adsabs.harvard.edu/abs/2014ApJ...785...14D/abstract}.
\end{barticle}
\endbibitem

\bibitem[\protect\citeauthoryear{Druckm{\"{u}}ller, Ru{\v{s}}in, and
  Minarovjech}{2006}]{Druckmuller2006}
\begin{barticle}
\bauthor{\bsnm{Druckm{\"{u}}ller}, \binits{M.}},
\bauthor{\bsnm{Ru{\v{s}}in}, \binits{V.}},
\bauthor{\bsnm{Minarovjech}, \binits{M.}}:
\byear{2006},
\batitle{{A new numerical method of total solar eclipse photography
  processing}}.
\bjtitle{Contrib. Astron. Obs. Skaln. Pleso}
\bvolume{36},
\bfpage{131}.
\burl{https://ui.adsabs.harvard.edu/abs/2006CoSka..36..131D/abstract}.
\end{barticle}
\endbibitem

\bibitem[\protect\citeauthoryear{Druckm{\"{u}}llerov{\'{a}}, Morgan, and
  Habbal}{2011}]{Druckmullerova2011}
\begin{barticle}
\bauthor{\bsnm{Druckm{\"{u}}llerov{\'{a}}}, \binits{H.}},
\bauthor{\bsnm{Morgan}, \binits{H.}},
\bauthor{\bsnm{Habbal}, \binits{S.R.}}:
\byear{2011},
\batitle{{Enhancing coronal structures with the fourier
  normalizing-radial-graded filter}}.
\bjtitle{Astrophys. J.}
\bvolume{737},
\bfpage{88}.
\doiurl{https://doi.org/10.1088/0004-637X/737/2/88}.
\end{barticle}
\endbibitem

\bibitem[\protect\citeauthoryear{Eddy}{1989}]{Eddy1989}
\begin{barticle}
\bauthor{\bsnm{Eddy}, \binits{J.A.}}:
\byear{1989},
\batitle{{Gordon Newkirk's Contributions to Coronal Studies}}.
\bjtitle{Highlights Astron.}
\bvolume{8},
\bfpage{503}.
\doiurl{https://doi.org/10.1017/s1539299600008182}.
\burl{https://ui.adsabs.harvard.edu/abs/1989HiA.....8..503E/abstract}.
\end{barticle}
\endbibitem

\bibitem[\protect\citeauthoryear{Freeland and Bentley}{2004}]{Freeland2004}
\begin{bchapter}
\bauthor{\bsnm{Freeland}, \binits{S.L.}},
\bauthor{\bsnm{Bentley}, \binits{R.D.}}:
\byear{2004},
\bctitle{{SolarSoft}}.
In: \bbtitle{Encycl. Astron. Astrophys.},
\bedition{1st} edn.,
\bpublisher{IOP Publishing Ltd}.
\doiurl{https://doi.org/10.1888/0333750888/3390}.
\end{bchapter}
\endbibitem

\bibitem[\protect\citeauthoryear{French et~al.}{2019}]{French2019}
\begin{barticle}
\bauthor{\bsnm{French}, \binits{R.J.}},
\bauthor{\bsnm{Judge}, \binits{P.G.}},
\bauthor{\bsnm{Matthews}, \binits{S.A.}},
\bauthor{\bparticle{van} \bsnm{Driel-Gesztelyi}, \binits{L.}}:
\byear{2019},
\batitle{{Spectropolarimetric Insight into Plasma Sheet Dynamics of a Solar
  Flare}}.
\bjtitle{Astrophys. J.}
\bvolume{887},
\bfpage{L34}.
\doiurl{https://doi.org/10.3847/2041-8213/AB5D34}.
\end{barticle}
\endbibitem

\bibitem[\protect\citeauthoryear{French et~al.}{2020}]{French2020}
\begin{barticle}
\bauthor{\bsnm{French}, \binits{R.J.}},
\bauthor{\bsnm{Matthews}, \binits{S.A.}},
\bauthor{\bparticle{van} \bsnm{Driel-Gesztelyi}, \binits{L.}},
\bauthor{\bsnm{Long}, \binits{D.M.}},
\bauthor{\bsnm{Judge}, \binits{P.G.}}:
\byear{2020},
\batitle{{Dynamics of Late-stage Reconnection in the 2017 September 10 Solar
  Flare}}.
\bjtitle{Astrophys. J.}
\bvolume{900},
\bfpage{192}.
\doiurl{https://doi.org/10.3847/1538-4357/ABA94B}.
\end{barticle}
\endbibitem

\bibitem[\protect\citeauthoryear{Gibson et~al.}{2016}]{Gibson2016b}
\begin{barticle}
\bauthor{\bsnm{Gibson}, \binits{S.E.}},
\bauthor{\bsnm{Kucera}, \binits{T.A.}},
\bauthor{\bsnm{White}, \binits{S.M.}},
\bauthor{\bsnm{Dove}, \binits{J.B.}},
\bauthor{\bsnm{Fan}, \binits{Y.}},
\bauthor{\bsnm{Forland}, \binits{B.C.}},
\bauthor{\bsnm{Rachmeler}, \binits{L.A.}},
\bauthor{\bsnm{Downs}, \binits{C.}},
\bauthor{\bsnm{Reeves}, \binits{K.K.}}:
\byear{2016},
\batitle{{FORWARD: A Toolset for Multiwavelength Coronal Magnetometry}}.
\bjtitle{Front. Astron. Sp. Sci.}
\bvolume{3},
\bfpage{8}.
\doiurl{https://doi.org/10.3389/fspas.2016.00008}.
\burl{https://ui.adsabs.harvard.edu/abs/2016FrASS...3....8G/abstract}.
\end{barticle}
\endbibitem

\bibitem[\protect\citeauthoryear{Gilly}{2022}]{Gilly2022}
\begin{botherref}
\oauthor{\bsnm{Gilly}, \binits{C.R.}}:
2022,
{Spectroscopic Analysis and Image Processing of the Optically-Thin Solar
  Corona}.
PhD thesis,
University of Colorado, Boulder.
\end{botherref}
\endbibitem

\bibitem[\protect\citeauthoryear{Gilly}{2024}]{Gilly2024}
\begin{botherref}
\oauthor{\bsnm{Gilly}, \binits{C.R.}}:
2024,
\textit{{Radial Histogram Equalization — sunkit{\_}image 0.5.2.dev60
  documentation}}.
\url{https://docs.sunpy.org/projects/sunkit-image/en/latest/generated/gallery/radial_histogram_equalization.html}.
\end{botherref}
\endbibitem

\bibitem[\protect\citeauthoryear{Gilly}{2025}]{dr_gilly_2025_17226415}
\begin{botherref}
\oauthor{\bsnm{Gilly}, \binits{C.R.}}:
2025,
\textit{{GillySpace27/IDL{\_}RHEF: v0.0.1 Initial Release}},
Zenodo.
\doiurl{https://doi.org/10.5281/zenodo.17226415}.
\url{https://doi.org/10.5281/zenodo.17226415}.
\end{botherref}
\endbibitem

\bibitem[\protect\citeauthoryear{Habbal et~al.}{2010}]{Habbal2010}
\begin{barticle}
\bauthor{\bsnm{Habbal}, \binits{S.R.}},
\bauthor{\bsnm{Druckm{\"{u}}ller}, \binits{M.}},
\bauthor{\bsnm{Morgan}, \binits{H.}},
\bauthor{\bsnm{Scholl}, \binits{I.}},
\bauthor{\bsnm{Ru{\v{s}}in}, \binits{V.}},
\bauthor{\bsnm{Daw}, \binits{A.}},
\bauthor{\bsnm{Johnson}, \binits{J.}},
\bauthor{\bsnm{Arndt}, \binits{M.}}:
\byear{2010},
\batitle{{Total solar eclipse observations of hot prominence shrouds}}.
\bjtitle{Astrophys. J.}
\bvolume{719},
\bfpage{1362}.
\doiurl{https://doi.org/10.1088/0004-637X/719/2/1362}.
\burl{https://ui.adsabs.harvard.edu/abs/2010ApJ...719.1362H/abstract}.
\end{barticle}
\endbibitem

\bibitem[\protect\citeauthoryear{Hummel}{1977}]{Hummel1977}
\begin{barticle}
\bauthor{\bsnm{Hummel}, \binits{R.}}:
\byear{1977},
\batitle{{Image enhancement by histogram transformation}}.
\bjtitle{Comput. Graph. Image Process.}
\bvolume{6},
\bfpage{184}.
\doiurl{https://doi.org/10.1016/S0146-664X(77)80011-7}.
\end{barticle}
\endbibitem

\bibitem[\protect\citeauthoryear{Ketcham}{1976}]{Ketcham1976}
\begin{barticle}
\bauthor{\bsnm{Ketcham}, \binits{D.J.}}:
\byear{1976},
\batitle{{Real-Time Image Enhancement Techniques}}.
\bjtitle{Image Process.}
\bvolume{0074},
\bfpage{120}.
\doiurl{https://doi.org/10.1117/12.954708}.
\burl{https://www.spiedigitallibrary.org/conference-proceedings-of-spie/0074/0000/Real-Time-Image-Enhancement-Techniques/10.1117/12.954708.full
  https://www.spiedigitallibrary.org/conference-proceedings-of-spie/0074/0000/Real-Time-Image-Enhancement-Techniques/10}.
\end{barticle}
\endbibitem

\bibitem[\protect\citeauthoryear{Kliem et~al.}{2013}]{Kliem2013}
\begin{barticle}
\bauthor{\bsnm{Kliem}, \binits{B.}},
\bauthor{\bsnm{Su}, \binits{Y.N.}},
\bauthor{\bsnm{{Van Ballegooijen}}, \binits{A.A.}},
\bauthor{\bsnm{Deluca}, \binits{E.E.}}:
\byear{2013},
\batitle{{Magnetohydrodynamic modeling of the solar eruption on 2010 April 8}}.
\bjtitle{Astrophys. J.}
\bvolume{779},
\bfpage{129}.
\doiurl{https://doi.org/10.1088/0004-637X/779/2/129}.
\burl{https://ui.adsabs.harvard.edu/abs/2013ApJ...779..129K/abstract}.
\end{barticle}
\endbibitem

\bibitem[\protect\citeauthoryear{Krista and Gallagher}{2009}]{Krista2009}
\begin{barticle}
\bauthor{\bsnm{Krista}, \binits{L.D.}},
\bauthor{\bsnm{Gallagher}, \binits{P.T.}}:
\byear{2009},
\batitle{{Automated coronal hole detection using local intensity thresholding
  techniques}}.
\bjtitle{Sol. Phys.}
\bvolume{256},
\bfpage{87}.
\doiurl{https://doi.org/10.1007/s11207-009-9357-2}.
\burl{https://ui.adsabs.harvard.edu/abs/2009SoPh..256...87K/abstract}.
\end{barticle}
\endbibitem

\bibitem[\protect\citeauthoryear{Lee et~al.}{2020}]{Lee2020}
\begin{barticle}
\bauthor{\bsnm{Lee}, \binits{J.-O.}},
\bauthor{\bsnm{Cho}, \binits{K.-S.}},
\bauthor{\bsnm{Lee}, \binits{K.-S.}},
\bauthor{\bsnm{Cho}, \binits{I.-H.}},
\bauthor{\bsnm{Lee}, \binits{J.}},
\bauthor{\bsnm{Miyashita}, \binits{Y.}},
\bauthor{\bsnm{Kim}, \binits{Y.-H.}},
\bauthor{\bsnm{Kim}, \binits{R.-S.}},
\bauthor{\bsnm{Jang}, \binits{S.}}:
\byear{2020},
\batitle{{Formation of Post-CME Blobs Observed by LASCO-C2 and K-Cor on 2017
  September 10}}.
\bjtitle{Astrophys. J.}
\bvolume{892},
\bfpage{129}.
\doiurl{https://doi.org/10.3847/1538-4357/ab799a}.
\end{barticle}
\endbibitem

\bibitem[\protect\citeauthoryear{Lemen et~al.}{2012}]{Lemen2012}
\begin{barticle}
\bauthor{\bsnm{Lemen}, \binits{J.R.}},
\bauthor{\bsnm{Title}, \binits{A.M.}},
\bauthor{\bsnm{Akin}, \binits{D.J.}},
\bauthor{\bsnm{Boerner}, \binits{P.F.}},
\bauthor{\bsnm{Chou}, \binits{C.}},
\bauthor{\bsnm{Drake}, \binits{J.F.}},
\bauthor{\bsnm{Duncan}, \binits{D.W.}},
\bauthor{\bsnm{Edwards}, \binits{C.G.}},
\bauthor{\bsnm{Friedlaender}, \binits{F.M.}},
\bauthor{\bsnm{Heyman}, \binits{G.F.}},
\bauthor{\bsnm{Hurlburt}, \binits{N.E.}},
\bauthor{\bsnm{Katz}, \binits{N.L.}},
\bauthor{\bsnm{Kushner}, \binits{G.D.}},
\bauthor{\bsnm{Levay}, \binits{M.}},
\bauthor{\bsnm{Lindgren}, \binits{R.W.}},
\bauthor{\bsnm{Mathur}, \binits{D.P.}},
\bauthor{\bsnm{McFeaters}, \binits{E.L.}},
\bauthor{\bsnm{Mitchell}, \binits{S.}},
\bauthor{\bsnm{Rehse}, \binits{R.A.}},
\bauthor{\bsnm{Schrijver}, \binits{C.}},
\bauthor{\bsnm{Springer}, \binits{L.A.}},
\bauthor{\bsnm{Stern}, \binits{R.A.}},
\bauthor{\bsnm{Tarbell}, \binits{T.D.}},
\bauthor{\bsnm{Wuelser}, \binits{J.-P.P.}},
\bauthor{\bsnm{Wolfson}, \binits{C.J.}},
\bauthor{\bsnm{Yanari}, \binits{C.}},
\bauthor{\bsnm{Bookbinder}, \binits{J.A.}},
\bauthor{\bsnm{Cheimets}, \binits{P.N.}},
\bauthor{\bsnm{Caldwell}, \binits{D.}},
\bauthor{\bsnm{Deluca}, \binits{E.E.}},
\bauthor{\bsnm{Gates}, \binits{R.}},
\bauthor{\bsnm{Golub}, \binits{L.}},
\bauthor{\bsnm{Park}, \binits{S.}},
\bauthor{\bsnm{Podgorski}, \binits{W.A.}},
\bauthor{\bsnm{Bush}, \binits{R.I.}},
\bauthor{\bsnm{Scherrer}, \binits{P.H.}},
\bauthor{\bsnm{Gummin}, \binits{M.A.}},
\bauthor{\bsnm{Smith}, \binits{P.}},
\bauthor{\bsnm{Auker}, \binits{G.}},
\bauthor{\bsnm{Jerram}, \binits{P.}},
\bauthor{\bsnm{Pool}, \binits{P.}},
\bauthor{\bsnm{Soufli}, \binits{R.}},
\bauthor{\bsnm{Windt}, \binits{D.L.}},
\bauthor{\bsnm{Beardsley}, \binits{S.}},
\bauthor{\bsnm{Clapp}, \binits{M.}},
\bauthor{\bsnm{Lang}, \binits{J.}},
\bauthor{\bsnm{Waltham}, \binits{N.}}:
\byear{2012},
\batitle{{The Atmospheric Imaging Assembly (AIA) on the Solar Dynamics
  Observatory (SDO)}}.
\bjtitle{Sol. Phys.}
\bvolume{275},
\bfpage{17}.
\bisbn{1120701197768}.
\doiurl{https://doi.org/10.1007/s11207-011-9776-8}.
\burl{https://ui.adsabs.harvard.edu/abs/2012SoPh..275...17L/abstract
  http://download.springer.com/static/pdf/28/art{\%}253A10.1007{\%}252Fs11207-011-9776-8.pdf?originUrl=http{\%}3A{\%}2F{\%}2Flink.springer.com{\%}2Farticle{\%}2F10.1007{\%}2Fs11207-011-9776-8{\%}5C{\&}token2=exp=1493774117{~}a}.
\end{barticle}
\endbibitem

\bibitem[\protect\citeauthoryear{L{\'{o}}pez-Portela
  et~al.}{2018}]{Lopez-Portela2018}
\begin{barticle}
\bauthor{\bsnm{L{\'{o}}pez-Portela}, \binits{C.}},
\bauthor{\bsnm{Panasenco}, \binits{O.}},
\bauthor{\bsnm{Blanco-Cano}, \binits{X.}},
\bauthor{\bsnm{Stenborg}, \binits{G.}}:
\byear{2018},
\batitle{{Deprojected Trajectory of Blobs in the Inner Corona}}.
\bjtitle{Sol. Phys.}
\bvolume{293},
\bfpage{1}.
\doiurl{https://doi.org/10.1007/s11207-018-1315-4}.
\burl{https://link.springer.com/article/10.1007/s11207-018-1315-4}.
\end{barticle}
\endbibitem

\bibitem[\protect\citeauthoryear{Ma et~al.}{2011}]{Ma2011}
\begin{barticle}
\bauthor{\bsnm{Ma}, \binits{S.}},
\bauthor{\bsnm{Raymond}, \binits{J.C.}},
\bauthor{\bsnm{Golub}, \binits{L.}},
\bauthor{\bsnm{Lin}, \binits{J.}},
\bauthor{\bsnm{Chen}, \binits{H.}},
\bauthor{\bsnm{Grigis}, \binits{P.}},
\bauthor{\bsnm{Testa}, \binits{P.}},
\bauthor{\bsnm{Long}, \binits{D.}}:
\byear{2011},
\batitle{{Observations and interpretation of a low coronal shock wave observed
  in the EUV by the SDO/AIA}}.
\bjtitle{Astrophys. J.}
\bvolume{738},
\bfpage{160}.
\doiurl{https://doi.org/10.1088/0004-637X/738/2/160}.
\burl{http://arxiv.org/abs/1106.6056
  http://dx.doi.org/10.1088/0004-637X/738/2/160}.
\end{barticle}
\endbibitem

\bibitem[\protect\citeauthoryear{Mart{\'{i}}nez-Galarce
  et~al.}{2013}]{Martinez-Galarce2013}
\begin{barticle}
\bauthor{\bsnm{Mart{\'{i}}nez-Galarce}, \binits{D.}},
\bauthor{\bsnm{Soufli}, \binits{R.}},
\bauthor{\bsnm{Windt}, \binits{D.L.}},
\bauthor{\bsnm{Bruner}, \binits{M.}},
\bauthor{\bsnm{Gullikson}, \binits{E.}},
\bauthor{\bsnm{Khatri}, \binits{S.}},
\bauthor{\bsnm{Spiller}, \binits{E.}},
\bauthor{\bsnm{Robinson}, \binits{J.C.}},
\bauthor{\bsnm{Baker}, \binits{S.}},
\bauthor{\bsnm{Prast}, \binits{E.}}:
\byear{2013},
\batitle{{Multisegmented, multilayer-coated mirrors for the Solar Ultraviolet
  Imager}}.
\bjtitle{Opt. Eng.}
\bvolume{52},
\bfpage{095102}.
\doiurl{https://doi.org/10.1117/1.oe.52.9.095102}.
\end{barticle}
\endbibitem

\bibitem[\protect\citeauthoryear{Masson et~al.}{2014}]{Masson2014}
\begin{barticle}
\bauthor{\bsnm{Masson}, \binits{S.}},
\bauthor{\bsnm{McCauley}, \binits{P.}},
\bauthor{\bsnm{Golub}, \binits{L.}},
\bauthor{\bsnm{Reeves}, \binits{K.K.}},
\bauthor{\bsnm{Deluca}, \binits{E.E.}}:
\byear{2014},
\batitle{{Dynamics of the transition corona}}.
\bjtitle{Astrophys. J.}
\bvolume{787},
\bfpage{145}.
\doiurl{https://doi.org/10.1088/0004-637X/787/2/145}.
\burl{https://ui.adsabs.harvard.edu/abs/2014ApJ...787..145M/abstract}.
\end{barticle}
\endbibitem

\bibitem[\protect\citeauthoryear{Miralles et~al.}{2001}]{Miralles2001}
\begin{barticle}
\bauthor{\bsnm{Miralles}, \binits{M.P.}},
\bauthor{\bsnm{Cranmer}, \binits{S.R.}},
\bauthor{\bsnm{Panasyuk}, \binits{A.V.}},
\bauthor{\bsnm{Romoli}, \binits{M.}},
\bauthor{\bsnm{Kohl}, \binits{J.L.}}:
\byear{2001},
\batitle{{Comparison of Empirical Models for Polar and Equatorial Coronal
  Holes}}.
\bjtitle{Astrophys. J.}
\bvolume{549},
\bfpage{L257}.
\doiurl{https://doi.org/10.1086/319166}.
\burl{https://ui.adsabs.harvard.edu/abs/2001ApJ...549L.257M/abstract}.
\end{barticle}
\endbibitem

\bibitem[\protect\citeauthoryear{Mishra, Srivastava, and
  Chen}{2020}]{Mishra2020}
\begin{barticle}
\bauthor{\bsnm{Mishra}, \binits{S.K.}},
\bauthor{\bsnm{Srivastava}, \binits{A.K.}},
\bauthor{\bsnm{Chen}, \binits{P.F.}}:
\byear{2020},
\batitle{{Large-Scale Vortex Motion and Multiple Plasmoid Ejection Due to
  Twisting Prominence Threads and Associated Reconnection}}.
\bjtitle{Sol. Phys.}
\bvolume{295},
\bfpage{1}.
\doiurl{https://doi.org/10.1007/s11207-020-01733-w}.
\burl{https://link.springer.com/article/10.1007/s11207-020-01733-w}.
\end{barticle}
\endbibitem

\bibitem[\protect\citeauthoryear{Morgan and
  Druckm{\"{u}}ller}{2014}]{Morgan2014}
\begin{barticle}
\bauthor{\bsnm{Morgan}, \binits{H.}},
\bauthor{\bsnm{Druckm{\"{u}}ller}, \binits{M.}}:
\byear{2014},
\batitle{{Multi-Scale Gaussian Normalization for Solar Image Processing}}.
\bjtitle{Sol. Phys.}
\bvolume{289},
\bfpage{2945}.
\doiurl{https://doi.org/10.1007/s11207-014-0523-9}.
\burl{https://link.springer.com/article/10.1007/s11207-014-0523-9}.
\end{barticle}
\endbibitem

\bibitem[\protect\citeauthoryear{Morgan and Habbal}{2010}]{Morgan2010}
\begin{barticle}
\bauthor{\bsnm{Morgan}, \binits{H.}},
\bauthor{\bsnm{Habbal}, \binits{S.R.}}:
\byear{2010},
\batitle{{A method for separating coronal mass ejections from the quiescent
  corona}}.
\bjtitle{Astrophys. J.}
\bvolume{711},
\bfpage{631}.
\doiurl{https://doi.org/10.1088/0004-637X/711/2/631}.
\burl{https://iopscience.iop.org/article/10.1088/0004-637X/711/2/631
  https://iopscience.iop.org/article/10.1088/0004-637X/711/2/631/meta}.
\end{barticle}
\endbibitem

\bibitem[\protect\citeauthoryear{Morgan and Kors{\'{o}}s}{2022}]{Morgan2022}
\begin{barticle}
\bauthor{\bsnm{Morgan}, \binits{H.}},
\bauthor{\bsnm{Kors{\'{o}}s}, \binits{M.B.}}:
\byear{2022},
\batitle{{Tracing the Magnetic Field Topology of the Quiet Corona Using
  Propagating Disturbances}}.
\bjtitle{Astrophys. J. Lett.}
\bvolume{933},
\bfpage{L27}.
\doiurl{https://doi.org/10.3847/2041-8213/AC7B7E}.
\end{barticle}
\endbibitem

\bibitem[\protect\citeauthoryear{Morgan, Byrne, and Habbal}{2012}]{Morgan2012}
\begin{barticle}
\bauthor{\bsnm{Morgan}, \binits{H.}},
\bauthor{\bsnm{Byrne}, \binits{J.P.}},
\bauthor{\bsnm{Habbal}, \binits{S.R.}}:
\byear{2012},
\batitle{{Automatically detecting and tracking coronal mass ejections. I.
  Separation of dynamic and quiescent components in coronagraph images}}.
\bjtitle{Astrophys. J.}
\bvolume{752},
\bfpage{144}.
\doiurl{https://doi.org/10.1088/0004-637X/752/2/144}.
\burl{https://ui.adsabs.harvard.edu/abs/2012ApJ...752..144M/abstract}.
\end{barticle}
\endbibitem

\bibitem[\protect\citeauthoryear{Morgan, Habbal, and
  Fineschi}{2007}]{Morgan2007}
\begin{barticle}
\bauthor{\bsnm{Morgan}, \binits{H.}},
\bauthor{\bsnm{Habbal}, \binits{S.R.}},
\bauthor{\bsnm{Fineschi}, \binits{S.}}:
\byear{2007},
\batitle{{Viewing Structure In Coronal Images}}.
\bjtitle{ESASP}
\bvolume{641},
\bfpage{15}.
\bisbn{92-9291-205-2}.
\burl{https://ui.adsabs.harvard.edu/abs/2007ESASP.641E..15M/abstract}.
\end{barticle}
\endbibitem

\bibitem[\protect\citeauthoryear{Morgan, Habbal, and Woo}{2006}]{Morgan2006}
\begin{barticle}
\bauthor{\bsnm{Morgan}, \binits{H.}},
\bauthor{\bsnm{Habbal}, \binits{S.R.}},
\bauthor{\bsnm{Woo}, \binits{R.}}:
\byear{2006},
\batitle{{The Depiction of Coronal Structure in White Light Images}}.
\bjtitle{Sol. Physics, Vol. 236, Issue 2, pp.263-272}
\bvolume{236},
\bfpage{263}.
\bisbn{1120700601136}.
\doiurl{https://doi.org/10.1007/s11207-006-0113-6}.
\burl{http://arxiv.org/abs/astro-ph/0602174
  http://dx.doi.org/10.1007/s11207-006-0113-6
  https://link.springer.com/article/10.1007/s11207-006-0113-6}.
\end{barticle}
\endbibitem

\bibitem[\protect\citeauthoryear{Morrill et~al.}{2006}]{Morrill2006}
\begin{barticle}
\bauthor{\bsnm{Morrill}, \binits{J.S.}},
\bauthor{\bsnm{Korendyke}, \binits{C.M.}},
\bauthor{\bsnm{Brueckner}, \binits{G.E.}},
\bauthor{\bsnm{Giovane}, \binits{F.}},
\bauthor{\bsnm{Howard}, \binits{R.A.}},
\bauthor{\bsnm{Koomen}, \binits{M.}},
\bauthor{\bsnm{Moses}, \binits{D.}},
\bauthor{\bsnm{Plunkett}, \binits{S.P.}},
\bauthor{\bsnm{Vourlidas}, \binits{A.}},
\bauthor{\bsnm{Esfandiari}, \binits{E.}},
\bauthor{\bsnm{Rich}, \binits{N.}},
\bauthor{\bsnm{Wang}, \binits{D.}},
\bauthor{\bsnm{Thernisien}, \binits{A.F.}},
\bauthor{\bsnm{Lamy}, \binits{P.}},
\bauthor{\bsnm{Llebaria}, \binits{A.}},
\bauthor{\bsnm{Biesecker}, \binits{D.}},
\bauthor{\bsnm{Michels}, \binits{D.}},
\bauthor{\bsnm{Gong}, \binits{Q.}},
\bauthor{\bsnm{Andrews}, \binits{M.}}:
\byear{2006},
\batitle{{Calibration of the SOHO/LASCO C3 white light coronagraph}}.
\bjtitle{Sol. Phys.}
\bvolume{233},
\bfpage{331}.
\doiurl{https://doi.org/10.1007/s11207-006-2058-1}.
\burl{https://ui.adsabs.harvard.edu/abs/2006SoPh..233..331M/abstract}.
\end{barticle}
\endbibitem

\bibitem[\protect\citeauthoryear{Morton, Tomczyk, and Pinto}{2016}]{Morton2016}
\begin{barticle}
\bauthor{\bsnm{Morton}, \binits{R.J.}},
\bauthor{\bsnm{Tomczyk}, \binits{S.}},
\bauthor{\bsnm{Pinto}, \binits{R.F.}}:
\byear{2016},
\batitle{{A global view of velocity fluctuations in the corona below 1.3
  {\$}R{\_}s{\$} with CoMP}}.
\bjtitle{Astrophys. J.}
\bvolume{828},
\bfpage{89}.
\doiurl{https://doi.org/10.3847/0004-637X/828/2/89}.
\burl{http://arxiv.org/abs/1608.01831
  http://dx.doi.org/10.3847/0004-637X/828/2/89
  https://iopscience.iop.org/article/10.3847/0004-637X/828/2/89
  https://iopscience.iop.org/article/10.3847/0004-637X/828/2/89/meta}.
\end{barticle}
\endbibitem

\bibitem[\protect\citeauthoryear{Mostafavi et~al.}{2022}]{Mostafavi2022}
\begin{barticle}
\bauthor{\bsnm{Mostafavi}, \binits{P.}},
\bauthor{\bsnm{Merkin}, \binits{V.G.}},
\bauthor{\bsnm{Provornikova}, \binits{E.}},
\bauthor{\bsnm{Sorathia}, \binits{K.}},
\bauthor{\bsnm{Arge}, \binits{C.N.}},
\bauthor{\bsnm{Garretson}, \binits{J.}}:
\byear{2022},
\batitle{{High-resolution Simulations of the Inner Heliosphere in Search of the
  Kelvin–Helmholtz Waves}}.
\bjtitle{Astrophys. J.}
\bvolume{925},
\bfpage{181}.
\doiurl{https://doi.org/10.3847/1538-4357/ac3fb4}.
\burl{https://ui.adsabs.harvard.edu/abs/2022ApJ...925..181M/abstract}.
\end{barticle}
\endbibitem

\bibitem[\protect\citeauthoryear{Newkirk}{1967}]{Newkirk1967}
\begin{barticle}
\bauthor{\bsnm{Newkirk}, \binits{G.}}:
\byear{1967},
\batitle{{Structure of the Solar Corona}}.
\bjtitle{Annu. Rev. Astron. Astrophys.}
\bvolume{5},
\bfpage{213}.
\doiurl{https://doi.org/10.1146/annurev.aa.05.090167.001241}.
\burl{https://ui.adsabs.harvard.edu/abs/1967ARA{\%}26A...5..213N/abstract}.
\end{barticle}
\endbibitem

\bibitem[\protect\citeauthoryear{O'Kane et~al.}{2019}]{Okane2019}
\begin{barticle}
\bauthor{\bsnm{O'Kane}, \binits{J.}},
\bauthor{\bsnm{Green}, \binits{L.}},
\bauthor{\bsnm{Long}, \binits{D.M.}},
\bauthor{\bsnm{Reid}, \binits{H.}}:
\byear{2019},
\batitle{{Stealth Coronal Mass Ejections from Active Regions}}.
\bjtitle{Astrophys. J.}
\bvolume{882},
\bfpage{85}.
\doiurl{https://doi.org/10.3847/1538-4357/AB371B}.
\end{barticle}
\endbibitem

\bibitem[\protect\citeauthoryear{Pant, Datta, and Banerjee}{2015}]{Pant2015}
\begin{barticle}
\bauthor{\bsnm{Pant}, \binits{V.}},
\bauthor{\bsnm{Datta}, \binits{A.}},
\bauthor{\bsnm{Banerjee}, \binits{D.}}:
\byear{2015},
\batitle{{Flows and waves in braided solar coronal magnetic structures}}.
\bjtitle{Astrophys. J. Lett.}
\bvolume{801},
\bfpage{L2}.
\doiurl{https://doi.org/10.1088/2041-8205/801/1/L2}.
\burl{https://ui.adsabs.harvard.edu/abs/2015ApJ...801L...2P/abstract}.
\end{barticle}
\endbibitem

\bibitem[\protect\citeauthoryear{Patel et~al.}{2022}]{Patel2022}
\begin{barticle}
\bauthor{\bsnm{Patel}, \binits{R.}},
\bauthor{\bsnm{Majumdar}, \binits{S.}},
\bauthor{\bsnm{Pant}, \binits{V.}},
\bauthor{\bsnm{Banerjee}, \binits{D.}}:
\byear{2022},
\batitle{{A Simple Radial Gradient Filter for Batch-Processing of Coronagraph
  Images}}.
\bjtitle{Sol. Phys.}
\bvolume{297},
\bfpage{27}.
\doiurl{https://doi.org/10.1007/s11207-022-01957-y}.
\burl{http://arxiv.org/abs/2201.13043
  http://dx.doi.org/10.1007/s11207-022-01957-y
  https://link.springer.com/10.1007/s11207-022-01957-y}.
\end{barticle}
\endbibitem

\bibitem[\protect\citeauthoryear{Pisano et~al.}{1998}]{Pisano1998}
\begin{barticle}
\bauthor{\bsnm{Pisano}, \binits{E.D.}},
\bauthor{\bsnm{Zong}, \binits{S.}},
\bauthor{\bsnm{Hemminger}, \binits{B.M.}},
\bauthor{\bsnm{DeLuca}, \binits{M.}},
\bauthor{\bsnm{Johnston}, \binits{R.E.}},
\bauthor{\bsnm{Muller}, \binits{K.}},
\bauthor{\bsnm{Braeuning}, \binits{M.P.}},
\bauthor{\bsnm{Pizer}, \binits{S.M.}}:
\byear{1998},
\batitle{{Contrast limited adaptive histogram equalization image processing to
  improve the detection of simulated spiculations in dense mammograms}}.
\bjtitle{J. Digit. Imaging}
\bvolume{11},
\bfpage{193}.
\doiurl{https://doi.org/10.1007/BF03178082}.
\burl{https://link.springer.com/article/10.1007/BF03178082}.
\end{barticle}
\endbibitem

\bibitem[\protect\citeauthoryear{Pizer}{1981}]{Pizer1981}
\begin{bchapter}
\bauthor{\bsnm{Pizer}, \binits{S.M.}}:
\byear{1981},
\bctitle{{Intensity Mappings: Linearization, Image-Based, User-Controlled}}.
In: \bbtitle{Adv. Disp. Technol. II}
\bseriesno{0271},
\bpublisher{SPIE},
\bfpage{21}.
\bisbn{9780892523030}.
\doiurl{https://doi.org/10.1117/12.931759}.
\burl{https://ui.adsabs.harvard.edu/abs/1981SPIE..271...21P/abstract}.
\end{bchapter}
\endbibitem

\bibitem[\protect\citeauthoryear{Pizer et~al.}{1987}]{Pizer1987}
\begin{barticle}
\bauthor{\bsnm{Pizer}, \binits{S.M.}},
\bauthor{\bsnm{Amburn}, \binits{E.P.}},
\bauthor{\bsnm{Austin}, \binits{J.D.}},
\bauthor{\bsnm{Cromartie}, \binits{R.}},
\bauthor{\bsnm{Geselowitz}, \binits{A.}},
\bauthor{\bsnm{Greer}, \binits{T.}},
\bauthor{\bsnm{{ter Haar Romeny}}, \binits{B.}},
\bauthor{\bsnm{Zimmerman}, \binits{J.B.}},
\bauthor{\bsnm{Zuiderveld}, \binits{K.}}:
\byear{1987},
\batitle{{Adaptive Histogram Equalization and Its Variations.}}
\bjtitle{Comput. Vision, Graph. Image Process.}
\bvolume{39},
\bfpage{355}.
\doiurl{https://doi.org/10.1016/S0734-189X(87)80186-X}.
\end{barticle}
\endbibitem

\bibitem[\protect\citeauthoryear{P{\"{o}}ssel}{2020}]{Possel2020}
\begin{barticle}
\bauthor{\bsnm{P{\"{o}}ssel}, \binits{M.}}:
\byear{2020},
\batitle{{A BEGINNER'S GUIDE TO WORKING WITH ASTRONOMICAL DATA}}.
\bjtitle{Open J. Astrophys.}
\bvolume{3},
\bfpage{2}.
\doiurl{https://doi.org/10.21105/astro.1905.13189}.
\burl{https://ui.adsabs.harvard.edu/abs/2020OJAp....3E...2P/abstract}.
\end{barticle}
\endbibitem

\bibitem[\protect\citeauthoryear{Provornikova et~al.}{2024}]{Provornikova2024}
\begin{barticle}
\bauthor{\bsnm{Provornikova}, \binits{E.}},
\bauthor{\bsnm{Merkin}, \binits{V.G.}},
\bauthor{\bsnm{Vourlidas}, \binits{A.}},
\bauthor{\bsnm{Malanushenko}, \binits{A.}},
\bauthor{\bsnm{Gibson}, \binits{S.E.}},
\bauthor{\bsnm{Winter}, \binits{E.}},
\bauthor{\bsnm{Arge}, \binits{N.}}:
\byear{2024},
\batitle{{MHD modeling of a geoeffective interplanetary CME with the magnetic
  topology informed by in-situ observations}}.
\bjtitle{ApJ}
\bvolume{977},
\bfpage{106}.
\doiurl{https://doi.org/10.3847/1538-4357/AD83B1}.
\burl{https://ui.adsabs.harvard.edu/abs/2024ApJ...977..106P/abstract
  http://arxiv.org/abs/2405.13069}.
\end{barticle}
\endbibitem

\bibitem[\protect\citeauthoryear{Qiang et~al.}{2020}]{Qiang2020}
\begin{barticle}
\bauthor{\bsnm{Qiang}, \binits{Z.}},
\bauthor{\bsnm{Bai}, \binits{X.}},
\bauthor{\bsnm{Ji}, \binits{K.}},
\bauthor{\bsnm{Liu}, \binits{H.}},
\bauthor{\bsnm{Shang}, \binits{Z.}}:
\byear{2020},
\batitle{{Enhancing coronal structures with radial local multi-scale filter}}.
\bjtitle{New Astron.}
\bvolume{79},
\bfpage{101383}.
\doiurl{https://doi.org/10.1016/J.NEWAST.2020.101383}.
\end{barticle}
\endbibitem

\bibitem[\protect\citeauthoryear{Ruminska et~al.}{2022}]{Ruminska2022}
\begin{barticle}
\bauthor{\bsnm{Ruminska}, \binits{A.}},
\bauthor{\bsnm{Bak-Strdlicka}, \binits{U.}},
\bauthor{\bsnm{Gibson}, \binits{S.E.}},
\bauthor{\bsnm{Fan}, \binits{Y.}}:
\byear{2022},
\batitle{{Coronal Cavities in CoMP Observations}}.
\bjtitle{Astrophys. J.}
\bvolume{926},
\bfpage{146}.
\doiurl{https://doi.org/10.3847/1538-4357/ac469c}.
\end{barticle}
\endbibitem

\bibitem[\protect\citeauthoryear{Seaton and Darnel}{2018}]{Seaton2018a}
\begin{barticle}
\bauthor{\bsnm{Seaton}, \binits{D.B.}},
\bauthor{\bsnm{Darnel}, \binits{J.M.}}:
\byear{2018},
\batitle{{Observations of an Eruptive Solar Flare in the Extended EUV Solar
  Corona}}.
\bjtitle{Astrophys. J. Lett.}
\bvolume{852},
\bfpage{L9}.
\doiurl{https://doi.org/10.3847/2041-8213/aaa28e}.
\burl{https://ui.adsabs.harvard.edu/abs/2018ApJ...852L...9S/abstract}.
\end{barticle}
\endbibitem

\bibitem[\protect\citeauthoryear{Seaton et~al.}{2023}]{Seaton2023}
\begin{barticle}
\bauthor{\bsnm{Seaton}, \binits{D.B.}},
\bauthor{\bsnm{Berghmans}, \binits{D.}},
\bauthor{\bsnm{Bloomfield}, \binits{D.S.}},
\bauthor{\bsnm{{De Groof}}, \binits{A.}},
\bauthor{\bsnm{D'Huys}, \binits{E.}},
\bauthor{\bsnm{Nicula}, \binits{B.}},
\bauthor{\bsnm{Rachmeler}, \binits{L.A.}},
\bauthor{\bsnm{West}, \binits{M.J.}}:
\byear{2023},
\batitle{{The SWAP Filter: A Simple Azimuthally Varying Radial Filter for
  Wide-Field EUV Solar Images}}.
\bjtitle{Sol. Phys.}
\bvolume{298}.
\doiurl{https://doi.org/10.1007/s11207-023-02183-w}.
\burl{http://dx.doi.org/10.1007/s11207-023-02183-w}.
\end{barticle}
\endbibitem

\bibitem[\protect\citeauthoryear{{Sheeley, Jr.} et~al.}{1997}]{SheeleyJr.1997}
\begin{barticle}
\bauthor{\bsnm{{Sheeley, Jr.}}, \binits{N.R.}},
\bauthor{\bsnm{Wang}, \binits{Y.}},
\bauthor{\bsnm{Hawley}, \binits{S.H.}},
\bauthor{\bsnm{Brueckner}, \binits{G.E.}},
\bauthor{\bsnm{Dere}, \binits{K.P.}},
\bauthor{\bsnm{Howard}, \binits{R.A.}},
\bauthor{\bsnm{Koomen}, \binits{M.J.}},
\bauthor{\bsnm{Korendyke}, \binits{C.M.}},
\bauthor{\bsnm{Michels}, \binits{D.J.}},
\bauthor{\bsnm{Paswaters}, \binits{S.E.}},
\bauthor{\bsnm{Socker}, \binits{D.G.}},
\bauthor{\bsnm{{St. Cyr}}, \binits{O.C.}},
\bauthor{\bsnm{Wang}, \binits{D.}},
\bauthor{\bsnm{Lamy}, \binits{P.L.}},
\bauthor{\bsnm{Llebaria}, \binits{A.}},
\bauthor{\bsnm{Schwenn}, \binits{R.}},
\bauthor{\bsnm{Simnett}, \binits{G.M.}},
\bauthor{\bsnm{Plunkett}, \binits{S.}},
\bauthor{\bsnm{Biesecker}, \binits{D.A.}}:
\byear{1997},
\batitle{{Measurements of Flow Speeds in the Corona Between 2 and 30 Rs}}.
\bjtitle{Astrophys. J.}
\bvolume{484},
\bfpage{472}.
\doiurl{https://doi.org/10.1086/304338}.
\burl{https://ui.adsabs.harvard.edu/abs/1997ApJ...484..472S/abstract}.
\end{barticle}
\endbibitem

\bibitem[\protect\citeauthoryear{Stenborg and Cobelli}{2003}]{Stenborg2003}
\begin{barticle}
\bauthor{\bsnm{Stenborg}, \binits{G.}},
\bauthor{\bsnm{Cobelli}, \binits{P.J.}}:
\byear{2003},
\batitle{{A wavelet packets equalization technique to reveal the multiple
  spatial-scale nature of coronal structures}}.
\bjtitle{Astron. Astrophys.}
\bvolume{398},
\bfpage{1185}.
\doiurl{https://doi.org/10.1051/0004-6361:20021687}.
\burl{http://dx.doi.org/10.1051/0004-6361:20021687}.
\end{barticle}
\endbibitem

\bibitem[\protect\citeauthoryear{Su et~al.}{2015}]{Su2015}
\begin{barticle}
\bauthor{\bsnm{Su}, \binits{Y.}},
\bauthor{\bsnm{Ballegooijen}, \binits{A.V.}},
\bauthor{\bsnm{McCauley}, \binits{P.}},
\bauthor{\bsnm{Ji}, \binits{H.}},
\bauthor{\bsnm{Reeves}, \binits{K.K.}},
\bauthor{\bsnm{Deluca}, \binits{E.E.}}:
\byear{2015},
\batitle{{MAGNETIC STRUCTURE and DYNAMICS of the ERUPTING SOLAR POLAR CROWN
  PROMINENCE on 2012 MARCH 12}}.
\bjtitle{Astrophys. J.}
\bvolume{807},
\bfpage{144}.
\doiurl{https://doi.org/10.1088/0004-637X/807/2/144}.
\burl{https://ui.adsabs.harvard.edu/abs/2015ApJ...807..144S/abstract}.
\end{barticle}
\endbibitem

\bibitem[\protect\citeauthoryear{Tadikonda et~al.}{2019}]{Tadikonda2019b}
\begin{barticle}
\bauthor{\bsnm{Tadikonda}, \binits{S.K.}},
\bauthor{\bsnm{Freesland}, \binits{D.C.}},
\bauthor{\bsnm{Minor}, \binits{R.R.}},
\bauthor{\bsnm{Seaton}, \binits{D.B.}},
\bauthor{\bsnm{Comeyne}, \binits{G.J.}},
\bauthor{\bsnm{Krimchansky}, \binits{A.}}:
\byear{2019},
\batitle{{Coronal Imaging with the Solar UltraViolet Imager}}.
\bjtitle{Sol. Phys.}
\bvolume{294},
\bfpage{28}.
\doiurl{https://doi.org/10.1007/s11207-019-1411-0}.
\burl{http://arxiv.org/abs/1901.08531
  http://dx.doi.org/10.1007/s11207-019-1411-0}.
\end{barticle}
\endbibitem

\bibitem[\protect\citeauthoryear{Telloni, Giordano, and
  Antonucci}{2019}]{Telloni2019}
\begin{barticle}
\bauthor{\bsnm{Telloni}, \binits{D.}},
\bauthor{\bsnm{Giordano}, \binits{S.}},
\bauthor{\bsnm{Antonucci}, \binits{E.}}:
\byear{2019},
\batitle{{On the Fast Solar Wind Heating and Acceleration Processes: A
  Statistical Study Based on the UVCS Survey Data}}.
\bjtitle{Astrophys. J. Lett.}
\bvolume{881},
\bfpage{L36}.
\doiurl{https://doi.org/10.3847/2041-8213/ab3731}.
\burl{https://ui.adsabs.harvard.edu/abs/2019ApJ...881L..36T/abstract}.
\end{barticle}
\endbibitem

\bibitem[\protect\citeauthoryear{Tomczyk et~al.}{2022}]{Tomczyk2022}
\begin{barticle}
\bauthor{\bsnm{Tomczyk}, \binits{S.}},
\bauthor{\bsnm{Burkepile}, \binits{J.}},
\bauthor{\bparticle{de} \bsnm{Wijn}, \binits{A.}},
\bauthor{\bsnm{Gibson}, \binits{S.E.}},
\bauthor{\bsnm{Gilbert}, \binits{H.R.}},
\bauthor{\bsnm{Landi}, \binits{E.}},
\bauthor{\bsnm{Lin}, \binits{H.}},
\bauthor{\bsnm{DeLuca}, \binits{E.}},
\bauthor{\bsnm{{Martinez Pillet}}, \binits{V.}},
\bauthor{\bsnm{Zhang}, \binits{J.}},
\bauthor{\bsnm{Laursen}, \binits{K.}},
\bauthor{\bsnm{Tomczyk}, \binits{S.}},
\bauthor{\bsnm{Burkepile}, \binits{J.}},
\bauthor{\bparticle{de} \bsnm{Wijn}, \binits{A.}},
\bauthor{\bsnm{Gibson}, \binits{S.E.}},
\bauthor{\bsnm{Gilbert}, \binits{H.R.}},
\bauthor{\bsnm{Landi}, \binits{E.}},
\bauthor{\bsnm{Lin}, \binits{H.}},
\bauthor{\bsnm{DeLuca}, \binits{E.}},
\bauthor{\bsnm{{Martinez Pillet}}, \binits{V.}},
\bauthor{\bsnm{Zhang}, \binits{J.}},
\bauthor{\bsnm{Laursen}, \binits{K.}}:
\byear{2022},
\batitle{{The Coronal Solar Magnetism Observatory: Overview and Recent
  Progress}}.
\bjtitle{AGUFM}
\bvolume{2022},
\bfpage{SH22B}.
\burl{https://ui.adsabs.harvard.edu/abs/2022AGUFMSH22B..05T/abstract}.
\end{barticle}
\endbibitem

\bibitem[\protect\citeauthoryear{Virtanen et~al.}{2020}]{Virtanen2020}
\begin{barticle}
\bauthor{\bsnm{Virtanen}, \binits{P.}},
\bauthor{\bsnm{Gommers}, \binits{R.}},
\bauthor{\bsnm{Oliphant}, \binits{T.E.}},
\bauthor{\bsnm{Haberland}, \binits{M.}},
\bauthor{\bsnm{Reddy}, \binits{T.}},
\bauthor{\bsnm{Cournapeau}, \binits{D.}},
\bauthor{\bsnm{Burovski}, \binits{E.}},
\bauthor{\bsnm{Peterson}, \binits{P.}},
\bauthor{\bsnm{Weckesser}, \binits{W.}},
\bauthor{\bsnm{Bright}, \binits{J.}},
\bauthor{\bparticle{van~der} \bsnm{Walt}, \binits{S.J.}},
\bauthor{\bsnm{Brett}, \binits{M.}},
\bauthor{\bsnm{Wilson}, \binits{J.}},
\bauthor{\bsnm{Millman}, \binits{K.J.}},
\bauthor{\bsnm{Mayorov}, \binits{N.}},
\bauthor{\bsnm{Nelson}, \binits{A.R.J.}},
\bauthor{\bsnm{Jones}, \binits{E.}},
\bauthor{\bsnm{Kern}, \binits{R.}},
\bauthor{\bsnm{Larson}, \binits{E.}},
\bauthor{\bsnm{Carey}, \binits{C.J.}},
\bauthor{\bsnm{Polat}, \binits{I.}},
\bauthor{\bsnm{Feng}, \binits{Y.}},
\bauthor{\bsnm{Moore}, \binits{E.W.}},
\bauthor{\bsnm{VanderPlas}, \binits{J.}},
\bauthor{\bsnm{Laxalde}, \binits{D.}},
\bauthor{\bsnm{Perktold}, \binits{J.}},
\bauthor{\bsnm{Cimrman}, \binits{R.}},
\bauthor{\bsnm{Henriksen}, \binits{I.}},
\bauthor{\bsnm{Quintero}, \binits{E.A.}},
\bauthor{\bsnm{Harris}, \binits{C.R.}},
\bauthor{\bsnm{Archibald}, \binits{A.M.}},
\bauthor{\bsnm{Ribeiro}, \binits{A.H.}},
\bauthor{\bsnm{Pedregosa}, \binits{F.}},
\bauthor{\bparticle{van} \bsnm{Mulbregt}, \binits{P.}},
\bauthor{\bsnm{Vijaykumar}, \binits{A.}},
\bauthor{\bsnm{Bardelli}, \binits{A.P.}},
\bauthor{\bsnm{Rothberg}, \binits{A.}},
\bauthor{\bsnm{Hilboll}, \binits{A.}},
\bauthor{\bsnm{Kloeckner}, \binits{A.}},
\bauthor{\bsnm{Scopatz}, \binits{A.}},
\bauthor{\bsnm{Lee}, \binits{A.}},
\bauthor{\bsnm{Rokem}, \binits{A.}},
\bauthor{\bsnm{Woods}, \binits{C.N.}},
\bauthor{\bsnm{Fulton}, \binits{C.}},
\bauthor{\bsnm{Masson}, \binits{C.}},
\bauthor{\bsnm{H{\"{a}}ggstr{\"{o}}m}, \binits{C.}},
\bauthor{\bsnm{Fitzgerald}, \binits{C.}},
\bauthor{\bsnm{Nicholson}, \binits{D.A.}},
\bauthor{\bsnm{Hagen}, \binits{D.R.}},
\bauthor{\bsnm{Pasechnik}, \binits{D.V.}},
\bauthor{\bsnm{Olivetti}, \binits{E.}},
\bauthor{\bsnm{Martin}, \binits{E.}},
\bauthor{\bsnm{Wieser}, \binits{E.}},
\bauthor{\bsnm{Silva}, \binits{F.}},
\bauthor{\bsnm{Lenders}, \binits{F.}},
\bauthor{\bsnm{Wilhelm}, \binits{F.}},
\bauthor{\bsnm{Young}, \binits{G.}},
\bauthor{\bsnm{Price}, \binits{G.A.}},
\bauthor{\bsnm{Ingold}, \binits{G.L.}},
\bauthor{\bsnm{Allen}, \binits{G.E.}},
\bauthor{\bsnm{Lee}, \binits{G.R.}},
\bauthor{\bsnm{Audren}, \binits{H.}},
\bauthor{\bsnm{Probst}, \binits{I.}},
\bauthor{\bsnm{Dietrich}, \binits{J.P.}},
\bauthor{\bsnm{Silterra}, \binits{J.}},
\bauthor{\bsnm{Webber}, \binits{J.T.}},
\bauthor{\bsnm{Slavi{\v{c}}}, \binits{J.}},
\bauthor{\bsnm{Nothman}, \binits{J.}},
\bauthor{\bsnm{Buchner}, \binits{J.}},
\bauthor{\bsnm{Kulick}, \binits{J.}},
\bauthor{\bsnm{Sch{\"{o}}nberger}, \binits{J.L.}},
\bauthor{\bsnm{{de Miranda Cardoso}}, \binits{J.V.}},
\bauthor{\bsnm{Reimer}, \binits{J.}},
\bauthor{\bsnm{Harrington}, \binits{J.}},
\bauthor{\bsnm{Rodr{\'{i}}guez}, \binits{J.L.C.}},
\bauthor{\bsnm{Nunez-Iglesias}, \binits{J.}},
\bauthor{\bsnm{Kuczynski}, \binits{J.}},
\bauthor{\bsnm{Tritz}, \binits{K.}},
\bauthor{\bsnm{Thoma}, \binits{M.}},
\bauthor{\bsnm{Newville}, \binits{M.}},
\bauthor{\bsnm{K{\"{u}}mmerer}, \binits{M.}},
\bauthor{\bsnm{Bolingbroke}, \binits{M.}},
\bauthor{\bsnm{Tartre}, \binits{M.}},
\bauthor{\bsnm{Pak}, \binits{M.}},
\bauthor{\bsnm{Smith}, \binits{N.J.}},
\bauthor{\bsnm{Nowaczyk}, \binits{N.}},
\bauthor{\bsnm{Shebanov}, \binits{N.}},
\bauthor{\bsnm{Pavlyk}, \binits{O.}},
\bauthor{\bsnm{Brodtkorb}, \binits{P.A.}},
\bauthor{\bsnm{Lee}, \binits{P.}},
\bauthor{\bsnm{McGibbon}, \binits{R.T.}},
\bauthor{\bsnm{Feldbauer}, \binits{R.}},
\bauthor{\bsnm{Lewis}, \binits{S.}},
\bauthor{\bsnm{Tygier}, \binits{S.}},
\bauthor{\bsnm{Sievert}, \binits{S.}},
\bauthor{\bsnm{Vigna}, \binits{S.}},
\bauthor{\bsnm{Peterson}, \binits{S.}},
\bauthor{\bsnm{More}, \binits{S.}},
\bauthor{\bsnm{Pudlik}, \binits{T.}},
\bauthor{\bsnm{Oshima}, \binits{T.}},
\bauthor{\bsnm{Pingel}, \binits{T.J.}},
\bauthor{\bsnm{Robitaille}, \binits{T.P.}},
\bauthor{\bsnm{Spura}, \binits{T.}},
\bauthor{\bsnm{Jones}, \binits{T.R.}},
\bauthor{\bsnm{Cera}, \binits{T.}},
\bauthor{\bsnm{Leslie}, \binits{T.}},
\bauthor{\bsnm{Zito}, \binits{T.}},
\bauthor{\bsnm{Krauss}, \binits{T.}},
\bauthor{\bsnm{Upadhyay}, \binits{U.}},
\bauthor{\bsnm{Halchenko}, \binits{Y.O.}},
\bauthor{\bsnm{V{\'{a}}zquez-Baeza}, \binits{Y.}}:
\byear{2020},
\batitle{{SciPy 1.0: fundamental algorithms for scientific computing in
  Python}}.
\bjtitle{Nat. Methods}
\bvolume{17},
\bfpage{261}.
\doiurl{https://doi.org/10.1038/s41592-019-0686-2}.
\burl{https://ui.adsabs.harvard.edu/abs/2020NatMe..17..261V/abstract}.
\end{barticle}
\endbibitem

\bibitem[\protect\citeauthoryear{Vourlidas and Howard}{2006}]{Vourlidas2006}
\begin{barticle}
\bauthor{\bsnm{Vourlidas}, \binits{A.}},
\bauthor{\bsnm{Howard}, \binits{R.A.}}:
\byear{2006},
\batitle{{The Proper Treatment of Coronal Mass Ejection Brightness: A New
  Methodology and Implications for Observations}}.
\bjtitle{Astrophys. J.}
\bvolume{642},
\bfpage{1216}.
\doiurl{https://doi.org/10.1086/501122}.
\burl{https://ui.adsabs.harvard.edu/abs/2006ApJ...642.1216V/abstract}.
\end{barticle}
\endbibitem

\bibitem[\protect\citeauthoryear{Wang and Stenborg}{2010}]{Wang2010}
\begin{barticle}
\bauthor{\bsnm{Wang}, \binits{Y.M.}},
\bauthor{\bsnm{Stenborg}, \binits{G.}}:
\byear{2010},
\batitle{{Spinning motions in coronal cavities}}.
\bjtitle{Astrophys. J. Lett.}
\bvolume{719},
\bfpage{L181}.
\doiurl{https://doi.org/10.1088/2041-8205/719/2/L181}.
\burl{https://ui.adsabs.harvard.edu/abs/2010ApJ...719L.181W/abstract}.
\end{barticle}
\endbibitem

\bibitem[\protect\citeauthoryear{Weberg, Morton, and
  McLaughlin}{2018}]{Weberg2018}
\begin{barticle}
\bauthor{\bsnm{Weberg}, \binits{M.J.}},
\bauthor{\bsnm{Morton}, \binits{R.J.}},
\bauthor{\bsnm{McLaughlin}, \binits{J.A.}}:
\byear{2018},
\batitle{{An Automated Algorithm for Identifying and Tracking Transverse Waves
  in Solar Images}}.
\bjtitle{Astrophys. J.}
\bvolume{852},
\bfpage{57}.
\doiurl{https://doi.org/10.3847/1538-4357/aa9e4a}.
\burl{http://arxiv.org/abs/1807.04842
  http://dx.doi.org/10.3847/1538-4357/aa9e4a}.
\end{barticle}
\endbibitem

\bibitem[\protect\citeauthoryear{West et~al.}{2023}]{West2022}
\begin{barticle}
\bauthor{\bsnm{West}, \binits{M.J.}},
\bauthor{\bsnm{Seaton}, \binits{D.B.}},
\bauthor{\bsnm{Wexler}, \binits{D.B.}},
\bauthor{\bsnm{Raymond}, \binits{J.C.}},
\bauthor{\bsnm{{Del Zanna}}, \binits{G.}},
\bauthor{\bsnm{Rivera}, \binits{Y.J.}},
\bauthor{\bsnm{Kobelski}, \binits{A.R.}},
\bauthor{\bsnm{Chen}, \binits{B.}},
\bauthor{\bsnm{DeForest}, \binits{C.}},
\bauthor{\bsnm{Golub}, \binits{L.}},
\bauthor{\bsnm{Caspi}, \binits{A.}},
\bauthor{\bsnm{Gilly}, \binits{C.R.}},
\bauthor{\bsnm{Kooi}, \binits{J.E.}},
\bauthor{\bsnm{Meyer}, \binits{K.A.}},
\bauthor{\bsnm{Alterman}, \binits{B.L.}},
\bauthor{\bsnm{Alzate}, \binits{N.}},
\bauthor{\bsnm{Andretta}, \binits{V.}},
\bauthor{\bsnm{Auch{\`{e}}re}, \binits{F.}},
\bauthor{\bsnm{Banerjee}, \binits{D.}},
\bauthor{\bsnm{Berghmans}, \binits{D.}},
\bauthor{\bsnm{Chamberlin}, \binits{P.}},
\bauthor{\bsnm{Chitta}, \binits{L.P.}},
\bauthor{\bsnm{Downs}, \binits{C.}},
\bauthor{\bsnm{Giordano}, \binits{S.}},
\bauthor{\bsnm{Harra}, \binits{L.}},
\bauthor{\bsnm{Higginson}, \binits{A.}},
\bauthor{\bsnm{Howard}, \binits{R.A.}},
\bauthor{\bsnm{Kumar}, \binits{P.}},
\bauthor{\bsnm{Mason}, \binits{E.}},
\bauthor{\bsnm{Mason}, \binits{J.P.}},
\bauthor{\bsnm{Morton}, \binits{R.J.}},
\bauthor{\bsnm{Nykyri}, \binits{K.}},
\bauthor{\bsnm{Patel}, \binits{R.}},
\bauthor{\bsnm{Rachmeler}, \binits{L.}},
\bauthor{\bsnm{Reardon}, \binits{K.P.}},
\bauthor{\bsnm{Reeves}, \binits{K.K.}},
\bauthor{\bsnm{Savage}, \binits{S.}},
\bauthor{\bsnm{Thompson}, \binits{B.J.}},
\bauthor{\bsnm{{Van Kooten}}, \binits{S.J.}},
\bauthor{\bsnm{Viall}, \binits{N.M.}},
\bauthor{\bsnm{Vourlidas}, \binits{A.}},
\bauthor{\bsnm{Zhukov}, \binits{A.N.}}:
\byear{2023},
\batitle{{Defining the Middle Corona}}.
\bjtitle{Sol. Phys.}
\bvolume{298},
\bfpage{arXiv:2208.04485}.
\bisbn{2208.04485v1}.
\doiurl{https://doi.org/10.1007/s11207-023-02170-1}.
\burl{http://arxiv.org/abs/2208.04485}.
\end{barticle}
\endbibitem

\bibitem[\protect\citeauthoryear{Williams et~al.}{2020}]{Williams2020}
\begin{barticle}
\bauthor{\bsnm{Williams}, \binits{T.}},
\bauthor{\bsnm{Walsh}, \binits{R.W.}},
\bauthor{\bsnm{Winebarger}, \binits{A.R.}},
\bauthor{\bsnm{Brooks}, \binits{D.H.}},
\bauthor{\bsnm{Cirtain}, \binits{J.W.}},
\bauthor{\bsnm{{De Pontieu}}, \binits{B.}},
\bauthor{\bsnm{Golub}, \binits{L.}},
\bauthor{\bsnm{Kobayashi}, \binits{K.}},
\bauthor{\bsnm{McKenzie}, \binits{D.E.}},
\bauthor{\bsnm{Morton}, \binits{R.J.}},
\bauthor{\bsnm{Peter}, \binits{H.}},
\bauthor{\bsnm{Rachmeler}, \binits{L.A.}},
\bauthor{\bsnm{Savage}, \binits{S.L.}},
\bauthor{\bsnm{Testa}, \binits{P.}},
\bauthor{\bsnm{Tiwari}, \binits{S.K.}},
\bauthor{\bsnm{Warren}, \binits{H.P.}},
\bauthor{\bsnm{Watkinson}, \binits{B.J.}}:
\byear{2020},
\batitle{{Is the High-Resolution Coronal Imager Resolving Coronal Strands?
  Results from AR 12712}}.
\bjtitle{Astrophys. J.}
\bvolume{892},
\bfpage{134}.
\doiurl{https://doi.org/10.3847/1538-4357/ab6dcf}.
\burl{https://ui.adsabs.harvard.edu/abs/2020ApJ...892..134W/abstract}.
\end{barticle}
\endbibitem

\bibitem[\protect\citeauthoryear{Yadav, Maheshwari, and
  Agarwal}{2014}]{Yadav2014}
\begin{bchapter}
\bauthor{\bsnm{Yadav}, \binits{G.}},
\bauthor{\bsnm{Maheshwari}, \binits{S.}},
\bauthor{\bsnm{Agarwal}, \binits{A.}}:
\byear{2014},
\bctitle{{Contrast limited adaptive histogram equalization based enhancement
  for real time video system}}.
In: \bbtitle{Proc. 2014 Int. Conf. Adv. Comput. Commun. Informatics, ICACCI
  2014},
\bpublisher{Institute of Electrical and Electronics Engineers Inc.},
\bfpage{2392}.
\bisbn{9781479930791}.
\doiurl{https://doi.org/10.1109/ICACCI.2014.6968381}.
\end{bchapter}
\endbibitem

\bibitem[\protect\citeauthoryear{Zhang et~al.}{2019}]{Zhang2019a}
\begin{barticle}
\bauthor{\bsnm{Zhang}, \binits{B.}},
\bauthor{\bsnm{Sorathia}, \binits{K.A.}},
\bauthor{\bsnm{Lyon}, \binits{J.G.}},
\bauthor{\bsnm{Merkin}, \binits{V.G.}},
\bauthor{\bsnm{Garretson}, \binits{J.S.}},
\bauthor{\bsnm{Wiltberger}, \binits{M.}}:
\byear{2019},
\batitle{{GAMERA: A Three-dimensional Finite-volume MHD Solver for
  Non-orthogonal Curvilinear Geometries}}.
\bjtitle{Astrophys. J. Suppl. Ser.}
\bvolume{244},
\bfpage{20}.
\doiurl{https://doi.org/10.3847/1538-4365/ab3a4c}.
\burl{https://ui.adsabs.harvard.edu/abs/2019ApJS..244...20Z/abstract}.
\end{barticle}
\endbibitem

\bibitem[\protect\citeauthoryear{Zuiderveld}{1994}]{Zuiderveld1994}
\begin{bchapter}
\bauthor{\bsnm{Zuiderveld}, \binits{K.}}:
\byear{1994},
\bctitle{{Contrast Limited Adaptive Histogram Equalization}}.
In: \bbtitle{Graph. Gems},
\bpublisher{Academic Press},
\bfpage{474}.
\doiurl{https://doi.org/10.1016/b978-0-12-336156-1.50061-6}.
\end{bchapter}
\endbibitem

\end{thebibliography}

\label{sec:bibliography}
% \clearpage
% \vfill
\end{document}